\documentclass[twocolumn,footinbib,preprintnumbers,floatfix,aps,prl,10pt]{revtex4-2}
\usepackage{amsmath,amssymb,graphicx,booktabs,bm,psfrag,color,slashed,mathtools}
\usepackage{hyperref}

\usepackage[dvipsnames]{xcolor}

\newcommand{\spac}{{\hspace{0.3mm}}}
 \def\nb{\bar{n}}

\makeatletter 
    
\renewcommand\onecolumngrid{
\do@columngrid{one}{\@ne}%
\def\set@footnotewidth{\onecolumngrid}
\def\footnoterule{\kern-6pt\hrule width 1.5in\kern6pt}%
}

\renewcommand\twocolumngrid{
        \def\footnoterule{
        \dimen@\skip\footins\divide\dimen@\thr@@
        \kern-\dimen@\hrule width.5in\kern\dimen@}
        \do@columngrid{mlt}{\tw@}
}%

\makeatother    

\begin{document}

\preprint{CERN-TH-2026-304,MIT-CTP/6156}
\preprint{November 19, 2026}

\title{Concerning Factorization at Lepton Colliders}

\author{Patrick Hager$^a$}
\author{Dominik Schwienbacher$^b$}

\affiliation{${}^a$CERN, Theoretical Physics Department, CH-1211 Geneva 23, Switzerland\\
${}^b$MIT Center For Theoretical Physics -- A Leinweber Institute, Cambridge, MA 02139, USA}

\begin{abstract}
\vspace{-3mm} 
We uncover super-leading logarithms in non-global observables at lepton colliders. 
They arise from Glauber interactions involving colored partons produced by electromagnetic splittings of the incoming leptons. 
A three-loop analysis establishes the resulting factorization breaking, which we reconcile with the use of lepton PDFs below the soft veto scale. 
We conjecture that soft--collinear factorization fails for exclusive observables even in pure QED by identifying a genuine Glauber region in massless QED.
This reveals a far richer infrared structure than naively expected.
\end{abstract}

\maketitle

Predictions for jet observables at high-energy colliders rely fundamentally on factorization and color coherence. 
Factorization separates the hard scattering at short distances from low-energy dynamics~\cite{Collins:1985ue,Collins:1988ig}.
Soft--collinear factorization further separates radiation along the jet and beam directions from wide-angle soft radiation.
These separations allow for the corresponding contributions to be calculated and evolved separately, enabling systematically improvable predictions and the resummation of large logarithmic corrections.
Color coherence, on the other side, provides an essential simplification: wide-angle soft radiation is sensitive only to the total color charge of a jet of collinear partons, but does not resolve its internal structure.
A narrow jet then acts as a single source of soft radiation.

At hadron colliders, this coherent picture can fail for radiation from an incoming parton.
When a parton in an outgoing jet splits into further collinear partons (time-like kinematics), the amplitude factorizes into an amplitude for the parent process and a universal splitting amplitude which depends only on the participating collinear particles~\cite{Bern:1995ix,Catani:1998nv,Kosower:1999xi,Bern:2004cz,Feige:2014wja,Guan:2024hlf,Duhr:2025cye}.
Radiation from an incoming parton instead produces a collinear configuration containing both an incoming and an outgoing particle (space-like kinematics).
This distinction matters, as complex phases associated with Glauber gluon exchange distinguish incoming from outgoing particles.
Because of these phases, the splitting amplitude for an initial-state collinear emission is no longer universal: it depends on the color and kinematics of the other partons in the process, which breaks strict collinear factorization at the amplitude level~\cite{Catani:2011st,Henn:2024qjq,Duhr:2025lyg,Buccioni:2026mfg,splittings}.
Restrictions on radiation then obstruct the real--virtual cancellations, leaving behind traces of factorization breaking in the cross section.

In non-global observables, such as gaps-between-jets, where a veto restricts higher-energy radiation into the rapidity gap, this mechanism produces super-leading logarithms (SLLs)~\cite{Forshaw:2006fk,Forshaw:2008cq}; double-logarithmic contributions to otherwise single-logarithmic observables~\cite{Dasgupta:2001sh}.
The leading SLLs have been resummed to all orders~\cite{Becher:2021zkk,Becher:2023mtx,Boer:2023jsy,Boer:2023ljq,Boer:2024xzy,Banerjee:2025kkq} in a factorization framework based on soft-collinear effective theory (SCET)~\cite{Bauer:2001yt,Bauer:2002nz,Beneke:2002ph,Beneke:2002ni} along with phenomenological studies~\cite{Becher:2024nqc,Forshaw:2025bmo,Forshaw:2025fif,Dasgupta:2025cgl,Becher:2026uih,Schalch:2026xxk,Dasgupta:2026mlm}.
The underlying coherence violation, however, is not restricted to non-global observables~\cite{Banfi:2010xy,Forshaw:2021fxs}.
Globalness alone does not ensure cancellation: measuring radiation everywhere is not the same as summing inclusively over it.

Using color conservation, the relevant Glauber phases can be expressed through correlations between the two incoming color charges, proportional to $\bm{T}_a\cdot\bm{T}_b$.
The vanishing color charges of elementary leptons then eliminate these phases in the original leptonic channel.
However, the cancellation carries over to the full cross section only if it is preserved once the QED evolution is activated.

Photon radiation from a lepton, followed by a $\gamma\to q\bar{q}$ splitting, provides a suitable example.
When both beams undergo such branchings, the hard scattering involves two incoming colored partons, even though each beam-collinear configuration remains color neutral.
Subsequent Glauber interactions act on this enlarged color space and distinguish the incoming from the outgoing partons.
Replacing each set of daughters by its respective colorless parent therefore invokes color coherence whose validity is precisely at issue.
The absence of a Glauber phase in the original leptonic channel before the splitting does not establish its cancellation in the radiatively generated channels \emph{after} the splitting.

In this Letter, we establish the radiative generation of SLLs in gaps-between-jets cross sections at lepton colliders.
An explicit calculation verifies consistency with ordinary lepton-PDF factorization below the veto scale.
The additional coherence-violating evolution occurs between the hard and veto scales while lepton-mass logarithms remain part of the usual collinear evolution.

The distinction between incoming and outgoing particles is relevant even in pure QED. 
For space-like splittings, electromagnetic phases are encoded in generalized Wilson lines with different causal prescriptions.
Charge conservation alone is not enough to collapse these lines into a factor determined by the parent charge, so even a neutral photon can acquire a non-vanishing Wilson-line phase. The commutativity of Abelian charges offers no protection against these phases.

We further identify a genuine Glauber region in a five-loop topology in massless QED. A light-by-light fermion loop links three charged eikonal lines to the measured beam-collinear radiation through a Glauber photon, transferring the collinear measurement scale into the virtual soft subgraph. We expect that this contribution survives the sum over diagrams and cuts for suitable exclusive observables and thereby violates soft--collinear factorization.
These genuine modes have a direct correspondence in the method of regions~\cite{Beneke:1997zp,Pak:2010pt,Jantzen:2011nz,Jantzen:2012mw} as so-called ``hidden regions''~\cite{Heinrich:2021dbf,Ma:2023hrt,Gardi:2024axt,Chen:2026dnj,Gardi:2026mcx}. Together with the generalized Wilson line, this demonstrates a far richer infrared structure of QED than generally assumed.

\begin{figure}
    \centering
    \includegraphics[width=0.8\linewidth]{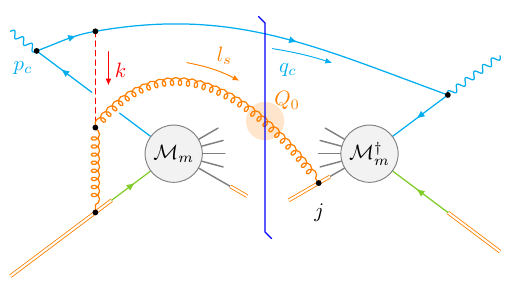} 
    \caption{Example diagram showing the first factorization-breaking contribution in photon--proton scattering. The collinear photon (blue) splits into a quark--anti-quark pair, and the genuine Glauber (red) transfers the measurement scale of the soft radiation (orange) into the collinear sub-diagram. The measurement of the soft radiation is depicted by an orange circle. The soft Wilson lines are denoted by orange double lines, and anti-collinear legs are depicted in green, while the hard amplitudes as well as outgoing radiation is shown in grey. }
    \label{fig:off_diag}
\end{figure}

We consider hard jets with transverse momenta of order $Q$, with radiation above $Q_0\ll Q$ vetoed in a rapidity interval $\Delta Y$ between them.
At leading power in $Q_0/Q\sim\lambda$, the hadronic cross section factorizes as~\cite{Becher:2021zkk,Becher:2023mtx}
\begin{equation}
\label{eq:factorization_theorem}
  \sigma(Q_0) =  
  \sum_{m=4}^{\infty} \langle\bm{\mathcal{H}}_m(\{n\}, Q, \mu) \otimes \bm{\mathcal{W}}_m(\{n\}, Q_0, \mu)\rangle\, .
\end{equation}
Here, $\bm{\mathcal{H}}_m$ are the hard functions with two incoming and $m-2$ outgoing partons, while $\bm{\mathcal{W}}_m$ contain the low-energy interactions and veto measurement.
The brackets contract open color and spin indices, and $\otimes$ includes integrations over directions and incoming momentum fractions. Glauber exchanges couple the soft and collinear sectors, so no further factorization is assumed in $\bm{\mathcal{W}}_m$.

The logarithms are resummed by evolving the hard functions from $\mu\sim Q$ to $\mu\sim Q_0$.
In the conventions of \cite{Becher:2025igg}, the one-loop QCD anomalous dimension~\cite{Becher:2009cu,Gardi:2009zv,Becher:2009qa} is
\begin{equation}
\label{eq:anomalous_dimension_soft_part}
   \bm\Gamma^{H} 
   = \frac{\alpha_s}{4\pi}\gamma^{\mathrm{cusp}}_0 \Bigl( \bm{\Gamma}^c
    \ln\frac{\mu^2}{Q^2} +\bm{V}^G \Bigr)
    + \frac{\alpha_s}{4\pi}\,\overline{\bm{\Gamma}} 
    + \frac{\alpha_s}{4\pi}\bm{\Gamma}^C \,.
\end{equation}
The operator $\overline{\bm{\Gamma}}$ describes wide-angle soft evolution, including virtual corrections and real emissions outside the veto region, while $\bm{\Gamma}^C$ contains the remaining initial-state collinear evolution. The two operators controlling the leading SLL mechanism are the cusp anomalous dimension $\bm{\Gamma}^c$ and the Glauber phase $\bm{V}^G$,
\begin{align} \label{eq:VGGammac}
    \bm{\Gamma}^c
    &= \sum_{i=1,2}\,\big[ C_i\,\bm{1} - \bm{T}_{i,L}\circ\bm{T}_{i,R}\,\delta(n_{l}-n_i) \big] \,,\\
    \bm{V}^G
    &= - 2i\pi\,\big( \bm{T}_{1,L}\cdot\bm{T}_{2,L} - \bm{T}_{1,R}\cdot\bm{T}_{2,R} \big)  \,. \nonumber 
\end{align}
Here, $C_i$ is the quadratic Casimir of the incoming parton $i$, and $L,R$ indicate action on the amplitude (left) and its conjugate (right).
The symbol $\circ$ denotes real emissions with the emitted gluon's color indices left open. The angular delta distribution fixes its direction $n_l$ to be collinear to the beam direction $n_i$.
The exact definitions of all appearing quantities can be found in~\cite{Becher:2025igg}.
Due to color conservation, it is always possible to transfer the final-state phases of the Glauber operator to initial-state color operators and commuting terms~\footnote{Note that the more intricate structure of the Glauber term discussed in~\cite{Becher:2026kbr} does not impact the general argument made here.}.

The logarithm in $\bm{\Gamma}^c$ generates double logarithms under scale evolution.
In coherent evolution, however, its real and virtual contributions cancel.
At hadron colliders, the Glauber phase $\bm{V}^G$ is non-vanishing and does not commute with $\bm{\Gamma}^c$. Two phase insertions together with one veto-sensitive soft evolution obstruct this cancellation, producing SLLs~\cite{Forshaw:2006fk,Forshaw:2008cq}\footnote{A pattern reminiscent of SLLs has also been observed at small-$x$ in the four-loop non-singlet quark splitting function~\cite{Davies:2022ofz,Gehrmann:2026qbl,Moch:2026qsw}.}.
For incoming elementary leptons, both operators vanish in the original leptonic channel and this effect is seemingly absent. 

At the very core of this assumption lies the intuition that only the color of the parent parton counts.
If each beam-collinear configuration could be replaced by its colorless parent lepton, the relevant QCD Glauber phases would directly cancel.
Therefore, one usually assumes the absence of these Glauber modes for leptonic collisions and the usual description of factorized low-energy dynamics in terms of soft functions and independent beam PDFs~\cite{Kuraev:1985hb,Blumlein:2004bs,Bertone:2019hks,Bertone:2022ktl,Schnubel:2025ejl,Frixione:2026rii} is then recovered~\footnote{Note that the use of electron PDFs is not necessary and does not impact the general argument of this Letter.}.

In the following, we show that this intuition is incorrect once QED effects are included.
When $\gamma\to q\bar{q}$ splittings generate colored incoming partons in both beams, they activate QCD Glauber phases that distinguish these partons from their outgoing collinear companions, resulting in factorization breaking.
The color neutrality of each beam-collinear system therefore does not protect the cross section from SLLs.

To isolate the mechanism, we first consider a photon scattering with a colored parton, as occurring in photon--proton collisions. 
The direct photon channel has no QCD Glauber phase in the anomalous dimension, since one incoming color charge vanishes.
A space-like $\gamma\to q\bar{q}$ splitting changes this situation: one daughter enters the hard scattering, while the other remains in the beam-collinear final state.
The hard process then has two incoming colored partons, and
the Glauber phase is activated.

We denote the corresponding splitting operator by $\bm{\Gamma}^C_{\gamma\to q\bar{q}}$, with the electromagnetic coupling $\alpha/(4\pi)$ factored out.
For one flavor and one choice of the active daughter, its color structure can be written as
\begin{equation}
    \bm{\Gamma}^C_{\gamma\to q\bar{q}}(z) \equiv 4 \mathcal{P}_{\gamma\to\bar{q}}(z)\spac\bm{\mathcal{C}}^{\phantom{\dagger}}_L\spac\bm{\mathcal{C}}_R^\dagger = 4\spac Q_q^2\spac (z^2+(1-z)^2)\spac\delta_{il}\delta_{jk} \,,
\end{equation}
where $Q_q$ is the electric charge of the quark in units of the elementary charge $e$, 
and the angular delta function fixing the emitted parton to the beam direction is implicit.
The splitting preserves total color, but maps into the quark--anti-quark color space, as indicated by the Kronecker deltas.
Glauber phases then distinguish the incoming daughter from its outgoing companion and prevent the reduction to the neutral parent photon.
Indeed, the associated renormalization group equations allow us to derive a consistency check of the poles in the low-energy matrix elements as in~\cite{Becher:2024kmk,Becher:2025igg}.
Writing these matrix elements schematically as
$\bm{\mathcal{W}}_m = \bm{\mathcal{I}}_m\otimes (f_1\, f_2)$,
the hard-function and PDF counterterms constrain the poles of the perturbative coefficient $\bm{\mathcal{I}}_m$.

The condition implies the pole structure of the low-energy matrix elements to be of the form
\begin{equation}\label{eq:Wm_poles_coll}
   \bm{\mathcal{I}}_m^{\mathrm{bare}}\ni
     \left( \frac{\alpha_s}{4\pi} \right)^2 \frac{\alpha}{4\pi}\frac{\bigl[\bm{\Gamma}^C_{\gamma\to q\bar{q}}\spac,\bm{V}^G\overline{\bm{\Gamma}}\bigr]}{12\varepsilon^3}\color{black} \,,
\end{equation}
where we focus on the relevant leading pole, with $d=4-2\varepsilon$.
Naively, this contribution is vanishing: as the $\gamma\to q\bar{q}$ splitting does not carry any color, the commutator with the soft emission and Glauber phase seems to evaluate to zero. 
However, the commutator is schematic, since the two terms act on different color spaces. Similar to the case of coherence-violating logarithms in $N$-jettiness~\cite{Becher:2026kbr}, the real-emission term in $\bm{\Gamma}^C_{\gamma\to q\bar{q}}$ breaks open the color space.
If $\bm{V}^G$ acts in the photon channel, the contribution vanishes.
In the other ordering, it acts in the channel with two colored incoming partons and distinguishes the incoming anti-quark and outgoing quark.
The two orderings therefore differ despite the vanishing total color charge, giving a non-zero operator structure.

\begin{figure}
    \centering
    \includegraphics[width=0.8\linewidth]{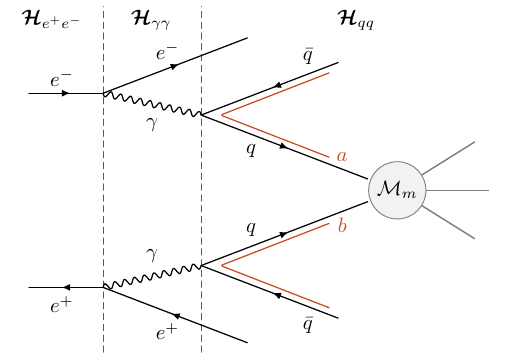} 
    \caption{Example splitting of an electron--positron pair into QCD initial states. The red line depicts color charge of the incoming partons, with color index $a,b$, while the indices on the hard function indicate the respective initial particles at the given stage.}
    \label{fig:coll}
\end{figure}

As the factorization breaking is now established, it is paramount that its corresponding counterpart appears within the low-energy matrix elements: if this was not the case, the definition of electron PDFs would change significantly, as it would depend on process-dependent information. 
Furthermore, this would no longer restrict double-logarithmic SLL evolution to the interval $(Q,Q_0)$.
As the electron mass $m_e$ provides a lower scale well below the intrinsic scale $\Lambda_{QCD}$, this would be highly problematic.
We now verify that this process-dependent color structure is
reproduced by the perturbative matching coefficient
$\bm{\mathcal I}_m$.\\

The required contributions arise from a Glauber interaction connecting the collinear splitting to a measured soft emission.
The collinear directions are given by light-like vectors $n_i$, such that $n_i^2=\bar{n}_i^2=0$ and $n_i\cdot\bar{n}_i=2$. 
We use a single set $n,\bar{n}$ for the incoming partons defined by ($n\cdot p_c,\bar{n}\cdot p_c, p_{c,\perp})\sim Q(\lambda^2,1,\lambda)$ for collinear direction and with $p_{\bar{c}}\sim Q(1,\lambda^2,\lambda)$ for anti-collinear momenta, with a small power-counting parameter $\lambda$. 
 
While soft radiation scales as $l_s\sim Q(\lambda,\lambda,\lambda)$, the Glaubers are distinct and scale as $k_g\sim Q(\lambda^2,\lambda,\lambda)$~\cite{Rothstein:2016bsq}\footnote{This is the scaling for collinear--soft Glauber, it is also possible to have anti-collinear--collinear Glaubers as well as anti-collinear--soft Glaubers which differ in scaling. Notably, the transverse momentum has to be compatible with both directions it mediates in between.}, thus being able to mediate between different sectors. 
The propagator of these modes is of the form $1/k^2=1/k_\perp^2$ and therefore they constitute instantaneous, Coulomb-like interactions and are, by construction, off-shell and can only interact with soft and collinear modes via their comparable transverse momentum scale.

In the massless partonic calculation, 
we can focus on a singular diagram responsible for factorization breaking, depicted in Fig.~\ref{fig:off_diag}. 
Any purely collinear loops and emissions are scaleless, as beam-collinear radiation is not resolved. 
Glauber exchange transfers the dependence on the measured soft momentum into the collinear subgraph, making the combined contribution nonzero, see also~\cite{Becher:2025igg} for further details.
Indeed, the calculation is similar to the one performed for the hadronic case.

At the splitting vertices, the replacement of the gluon by a photon changes the color factor $t_R^b t_L^a t_L^b$ into $Q_q^2\spac t_L^a$, while the Glauber and soft interactions remain QCD vertices. One then obtains
\begin{align}
    D_{\bar{q}/\gamma} &=  \alpha_s^2\alpha\frac{1}{2\spac\varepsilon^2} f^{ade} \sum_{j\neq1,2}\bm{T}_{2,L}^d \bm{T}_{j,R}^e \Bigl(\frac{\slashed{n}}{2}\Bigr)_{\beta\alpha} t^a_L 
        \\
    &\quad\times
    \int[d\Omega_l]
    \Bigl(\frac{n_j\cdot\nb}{n_j\cdot n_l\nb\cdot n_l} - \frac{n_j\cdot n}{n\cdot n_l\,n_j\cdot n_l}\Bigr)\Theta_{\mathrm{veto}}(n_l)
    \nonumber\\
    &\quad\times
    \mu^{6\varepsilon}
    \int\frac{dE_l}{(2\pi)^2}E_l^{-1-6\varepsilon}\,\theta(Q_0-E_l^T)
   \mathcal{P}_{\gamma\to \bar{q}}(z)
    \,.\label{app:eq:g_to_q_three_loop}\nonumber
\end{align}

Here, we omitted higher terms in $\varepsilon$ to allow for a cleaner notation, and $\Theta_{\mathrm{veto}}$ selects the gap, while $E_l^T$
is the transverse energy of the measured soft emission.
Together with the complex-conjugated counter piece and after summation over all legs $j$ this indeed reproduces~\eqref{eq:Wm_poles_coll}.
This verifies compatibility with ordinary PDF evolution for the displayed coherence-violating structure: the process-dependent contribution is accounted for completely by perturbative matching at the veto scale.

These results establish the corresponding mechanism for elementary lepton beams. 
The branchings $\ell\to\ell\gamma$ and $\gamma\to q\bar q$ generate colored incoming partons in both beams, as illustrated in Fig.~\ref{fig:coll}.
For a fixed incoming partonic channel, the perturbative coefficient is independent of whether the parton originates from a proton or a lepton. 
The generated channels therefore inherit the known non-zero partonic SLL contributions~\cite{Becher:2021zkk,Becher:2023mtx}.
The matching calculation above verifies compatibility of this mechanism with ordinary lepton PDFs at three-loop order.
For the case in which all four electromagnetic branchings occur between $Q$ and $Q_0$, leading-logarithmic power counting gives
\begin{align}
\delta \sigma_{\textrm{SLL}}&\sim\alpha_s^{n+3}\alpha^4\langle\bm{\mathcal{H}}_m \!\left(\bm{\Gamma}^c\right)^{n-r}\!\bm{V}^G\!\left(\bm{\Gamma}^c\right)^{r}\!\bm{V}^G\overline{\bm{\Gamma}}\spac\bm{\Gamma}_{\textrm{QED}}^4\rangle\nonumber\\
&\sim \hat{\sigma}_{\mathrm{Born}}\,\alpha^4 L^4 \alpha_s^{n+3} L^{2n+3}\,,
\end{align}
with $n\geq0$, $L=\ln(Q/Q_0)$, and $\hat\sigma_{\mathrm{Born}}$
denoting the Born cross section of the colored partonic channel.
The object $\bm{\Gamma}_{\textrm{QED}}^4$ schematically describes the necessary four QED splittings.
Beam-collinear branchings below the veto scale, which depend on lepton masses, are absorbed into the lepton PDFs~\cite{Bertone:2019hks,Han:2021kes,Garosi:2023bvq}, so this expression describes only one contribution and not the complete logarithmic structure.
The perturbatively generated colored components of the beams are sufficient to expose the failure of the protection by elementary, colorless leptons.

Although the overall effect is suppressed by powers of $\alpha$, it demonstrates that factorization breaking also occurs in leptonic collisions. Beyond its numerical impact, it implies the generalized factorization theorem~\eqref{eq:factorization_theorem} also for lepton colliders once QED evolution is included and highlights the importance of knowing the limits of the conventional factorization formulas.

While sufficiently inclusive observables are free of such effects~\cite{Bodwin:1984hc,Collins:1985ue,Collins:1988ig,Aybat:2008ct,Sterman:2022gyf,Sterman:2026vuh}, this does not imply that it is restricted to non-global observables. Also global observables such as $N$-jettiness feature coherence-violating logarithms~\cite{Banfi:2010xy,Forshaw:2021fxs,Banfi:2025mra,Becher:2026kbr}, and the previous analysis implies this also for lepton colliders.

\begin{figure}
    \centering
\includegraphics[width=0.8\linewidth]{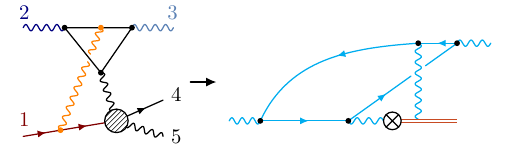} 
    \caption{Example matching of the factorization-breaking terms in QED from an amplitude point of view to SCET. The generalized Wilson line is depicted as a red double-line, effectively pushing the collinear mode attaching to it into a Glauber limit. This diagram shows the Abelian terms, which remain. The diagram to the left is adapted from~\cite{Buccioni:2026mfg}.}
    \label{fig:Wilson_lepton}
\end{figure}

The mechanism behind factorization breaking relies on three ingredients: space-like splittings, Glauber exchange, and a sufficiently exclusive measurement on (color-)charged final states. 
Indeed, all three requirements are present even in pure QED.
A concrete example is the space-like collinear limit of $e e\to e e \gamma$, with an outgoing electron collinear to the incoming one.
The incoming and outgoing electrons generate a residual Wilson-line combination of the form $[W^+_{i,e}]^\dagger W^-_{i,e}$ adapting the notation of~\cite{Becher:2026kbr,splittings}.
Both Wilson lines carry the electron charge and the superscript specifies the eikonal prescription.
Although the ordinary parent-charge Wilson line is trivial for the photon, this combination is not.
For a single emission, after factoring out the common charge and eikonal numerator, the residual denominator is
\begin{equation}
    \frac{1}{\bar{n}\cdot k-i0} - \frac{1}{\nb\cdot k+i0} = 2\pi i\,\delta(\bar{n}\cdot k)\,.
\end{equation}
Commutativity of the charges does not remove the difference between the causal prescriptions. Through interactions with the other charged legs, this residual Wilson-line factor introduces process dependence even when the parent in the reduced amplitude is a neutral photon.

The generalized Wilson-line formalism also reproduces the complete two-loop space-like splitting amplitudes in QCD~\cite{splittings}.
The fermion-loop contributions provide an explicit Abelian example, illustrated in Fig.~\ref{fig:Wilson_lepton}.
Their color factor reduces to $2\sum_f Q_f^4$, where $Q_f$ is the charge of the fermion in the loop.
Although nonzero at amplitude level, the corresponding factorization-violating terms cancel in the squared amplitude through the two-loop order~\cite{Forshaw:2012bi,Buccioni:2026mfg}.
However, at higher orders they contribute to the physical cross section. 

\begin{figure}
    \centering
    \includegraphics[width=0.8\linewidth]{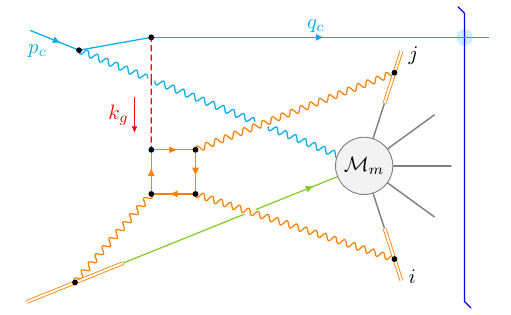} 
    \caption{Candidate genuine-Glauber contribution in massless QED, in the same color scheme as Fig.~\ref{fig:off_diag}. 
    A genuine Glauber photon carries the measurement of the collinear line, indicated by the blue circle, into the soft loops.}
    \label{fig:5_loop_qed}
\end{figure}

More generally, we conjecture that also genuine Glauber photons appear at higher orders. To illustrate this, consider a jettiness-like observable~\cite{Stewart:2010tn} with $\mathcal{T}_N\ll Q$, whose global nature makes it a particularly clean example. 
Even though real soft radiation of order of the collinear scale $\sqrt{Q\mathcal{T}_N}$ is excluded by the measurement, soft modes can contribute virtually~\cite{Becher:2026kbr}.

In the topology depicted in Fig.~\ref{fig:5_loop_qed}, a Glauber photon connects the measured collinear radiation to the virtual soft loops, transferring the scale-dependence from the measurement into the soft subgraph. This supplies a scale that would be absent in the isolated virtual soft integrals.
We explicitly identified a momentum configuration required for this Glauber exchange supporting the proposed mechanism and verified its existence with the Hidden Region Finder~\cite{Gardi:2026mcx}.
Such a contribution breaks factorization both kinematically and in its Lorentz structure, in contrast to the generalized collinear Wilson lines discussed above, indicating that factorization breaking persists at high loop orders even in pure QED.
The soft subgraph is precisely the light-by-light topology underlying the three-loop tripole contribution to the massless QED soft-photon current anticipated in~\cite{Dixon:2019lnw}. A massless-fermion box couples three charged eikonal lines to a fourth photon, generating correlations among their directions without non-Abelian gauge interactions.
In our diagram, this fourth photon carries the space-like Glauber momentum into the measured collinear sector.
The same topology connects the proposed correction to the QED soft current with a concrete mechanism for soft--collinear factorization breaking.
Additional contributions may also arise from double space-like splittings, provided the observable is sufficiently exclusive~\cite{Gaunt:2014ska,Zeng:2015iba, Duhr:2025lyg}.
The inclusion of electron-mass effects at lower scales is also an interesting aspect, which is beyond the scope of this work.
We anticipate that this mechanism produces a non-zero factorization-violating contribution to the exclusive cross section at five-loop order. 
Establishing its survival requires the complete sum over the relevant diagrams, which is beyond the scope of this work.

In this Letter, we have shown that lepton colliders are not immune to factorization breaking. 
Electromagnetic splittings generate colored incoming partons whose Glauber interactions produce super-leading logarithms, despite the color neutrality of the parent beams.
The underlying mechanism is not restricted to non-global observables but affects also global observables, similar to the hadronic case~\cite{Banfi:2025mra,Becher:2026kbr}.
The explicit matching calculation furthermore verifies compatibility of the coherence-violating contribution with ordinary lepton PDFs.

Although the radiatively generated QCD effects are suppressed by electromagnetic couplings, their conceptual implication is drastic: the elementary nature and color neutrality of lepton beams do not guarantee the cancellation of Glauber phases.
Space-like splittings require generalized Wilson lines, further challenging the conventional coherent picture for pure QED even for neutral photons.
In addition, genuine Glauber photon exchanges starting at higher loops are expected to yield additional factorization-breaking contributions.

It would be interesting to study the phenomenological implications of these effects in photon--photon collisions occurring in ultra-peripheral nucleon collisions at the LHC, and to explore the prospects of their observation at the upcoming FCC.
Here, also diffraction~\cite{Collins:1992cv,Fichet:2014uka,Frankfurt:2022jns,Lee:2025fml,Aretz:2026sjb} and Regge kinematics~\cite{Caola:2021izf,Falcioni:2021buo,Falcioni:2021dgr,Neill:2023jcd,Gao:2024qsg,Buccioni:2024gzo,Gao:2024fyz,Barcaro:2026dsd} will be relevant.
More broadly, our findings imply that QED at higher-loop orders proves to be much richer and more intricate than naively expected and many new phenomena remain to be uncovered.

\begin{acknowledgments}
{\em Acknowledgements --- \/} We would like to thank Thomas Becher, Federico Buccioni, and Robert Szafron for discussions and comments on the manuscript. DS was funded by the Swiss National Science Foundation Postdoc.mobility grant Nr.~P500-2\_239446.
\end{acknowledgments}

\bibliography{Lepton}

\begin{thebibliography}{90}%
\makeatletter
\providecommand \@ifxundefined [1]{%
 \@ifx{#1\undefined}
}%
\providecommand \@ifnum [1]{%
 \ifnum #1\expandafter \@firstoftwo
 \else \expandafter \@secondoftwo
 \fi
}%
\providecommand \@ifx [1]{%
 \ifx #1\expandafter \@firstoftwo
 \else \expandafter \@secondoftwo
 \fi
}%
\providecommand \natexlab [1]{#1}%
\providecommand \enquote  [1]{``#1''}%
\providecommand \bibnamefont  [1]{#1}%
\providecommand \bibfnamefont [1]{#1}%
\providecommand \citenamefont [1]{#1}%
\providecommand \href@noop [0]{\@secondoftwo}%
\providecommand \href [0]{\begingroup \@sanitize@url \@href}%
\providecommand \@href[1]{\@@startlink{#1}\@@href}%
\providecommand \@@href[1]{\endgroup#1\@@endlink}%
\providecommand \@sanitize@url [0]{\catcode `\\12\catcode `\$12\catcode `\&12\catcode `\#12\catcode `\^12\catcode `\_12\catcode `\%12\relax}%
\providecommand \@@startlink[1]{}%
\providecommand \@@endlink[0]{}%
\providecommand \url  [0]{\begingroup\@sanitize@url \@url }%
\providecommand \@url [1]{\endgroup\@href {#1}{\urlprefix }}%
\providecommand \urlprefix  [0]{URL }%
\providecommand \Eprint [0]{\href }%
\providecommand \doibase [0]{https://doi.org/}%
\providecommand \selectlanguage [0]{\@gobble}%
\providecommand \bibinfo  [0]{\@secondoftwo}%
\providecommand \bibfield  [0]{\@secondoftwo}%
\providecommand \translation [1]{[#1]}%
\providecommand \BibitemOpen [0]{}%
\providecommand \bibitemStop [0]{}%
\providecommand \bibitemNoStop [0]{.\EOS\space}%
\providecommand \EOS [0]{\spacefactor3000\relax}%
\providecommand \BibitemShut  [1]{\csname bibitem#1\endcsname}%
\let\auto@bib@innerbib\@empty
\bibitem [{\citenamefont {Collins}\ \emph {et~al.}(1985)\citenamefont {Collins}, \citenamefont {Soper},\ and\ \citenamefont {Sterman}}]{Collins:1985ue}%
  \BibitemOpen
  \bibfield  {author} {\bibinfo {author} {\bibfnamefont {J.~C.}\ \bibnamefont {Collins}}, \bibinfo {author} {\bibfnamefont {D.~E.}\ \bibnamefont {Soper}},\ and\ \bibinfo {author} {\bibfnamefont {G.~F.}\ \bibnamefont {Sterman}},\ }\bibfield  {title} {\bibinfo {title} {{Factorization for Short Distance Hadron - Hadron Scattering}},\ }\href {https://doi.org/10.1016/0550-3213(85)90565-6} {\bibfield  {journal} {\bibinfo  {journal} {Nucl. Phys. B}\ }\textbf {\bibinfo {volume} {261}},\ \bibinfo {pages} {104} (\bibinfo {year} {1985})}\BibitemShut {NoStop}%
\bibitem [{\citenamefont {Collins}\ \emph {et~al.}(1988)\citenamefont {Collins}, \citenamefont {Soper},\ and\ \citenamefont {Sterman}}]{Collins:1988ig}%
  \BibitemOpen
  \bibfield  {author} {\bibinfo {author} {\bibfnamefont {J.~C.}\ \bibnamefont {Collins}}, \bibinfo {author} {\bibfnamefont {D.~E.}\ \bibnamefont {Soper}},\ and\ \bibinfo {author} {\bibfnamefont {G.~F.}\ \bibnamefont {Sterman}},\ }\bibfield  {title} {\bibinfo {title} {{Soft Gluons and Factorization}},\ }\href {https://doi.org/10.1016/0550-3213(88)90130-7} {\bibfield  {journal} {\bibinfo  {journal} {Nucl. Phys. B}\ }\textbf {\bibinfo {volume} {308}},\ \bibinfo {pages} {833} (\bibinfo {year} {1988})}\BibitemShut {NoStop}%
\bibitem [{\citenamefont {Bern}\ and\ \citenamefont {Chalmers}(1995)}]{Bern:1995ix}%
  \BibitemOpen
  \bibfield  {author} {\bibinfo {author} {\bibfnamefont {Z.}~\bibnamefont {Bern}}\ and\ \bibinfo {author} {\bibfnamefont {G.}~\bibnamefont {Chalmers}},\ }\bibfield  {title} {\bibinfo {title} {{Factorization in one loop gauge theory}},\ }\href {https://doi.org/10.1016/0550-3213(95)00226-I} {\bibfield  {journal} {\bibinfo  {journal} {Nucl. Phys. B}\ }\textbf {\bibinfo {volume} {447}},\ \bibinfo {pages} {465} (\bibinfo {year} {1995})},\ \Eprint {https://arxiv.org/abs/hep-ph/9503236} {arXiv:hep-ph/9503236} \BibitemShut {NoStop}%
\bibitem [{\citenamefont {Catani}\ and\ \citenamefont {Grazzini}(1999)}]{Catani:1998nv}%
  \BibitemOpen
  \bibfield  {author} {\bibinfo {author} {\bibfnamefont {S.}~\bibnamefont {Catani}}\ and\ \bibinfo {author} {\bibfnamefont {M.}~\bibnamefont {Grazzini}},\ }\bibfield  {title} {\bibinfo {title} {{Collinear factorization and splitting functions for next-to-next-to-leading order QCD calculations}},\ }\href {https://doi.org/10.1016/S0370-2693(98)01513-5} {\bibfield  {journal} {\bibinfo  {journal} {Phys. Lett. B}\ }\textbf {\bibinfo {volume} {446}},\ \bibinfo {pages} {143} (\bibinfo {year} {1999})},\ \Eprint {https://arxiv.org/abs/hep-ph/9810389} {arXiv:hep-ph/9810389} \BibitemShut {NoStop}%
\bibitem [{\citenamefont {Kosower}(1999)}]{Kosower:1999xi}%
  \BibitemOpen
  \bibfield  {author} {\bibinfo {author} {\bibfnamefont {D.~A.}\ \bibnamefont {Kosower}},\ }\bibfield  {title} {\bibinfo {title} {{All order collinear behavior in gauge theories}},\ }\href {https://doi.org/10.1016/S0550-3213(99)00251-5} {\bibfield  {journal} {\bibinfo  {journal} {Nucl. Phys. B}\ }\textbf {\bibinfo {volume} {552}},\ \bibinfo {pages} {319} (\bibinfo {year} {1999})},\ \Eprint {https://arxiv.org/abs/hep-ph/9901201} {arXiv:hep-ph/9901201} \BibitemShut {NoStop}%
\bibitem [{\citenamefont {Bern}\ \emph {et~al.}(2004)\citenamefont {Bern}, \citenamefont {Dixon},\ and\ \citenamefont {Kosower}}]{Bern:2004cz}%
  \BibitemOpen
  \bibfield  {author} {\bibinfo {author} {\bibfnamefont {Z.}~\bibnamefont {Bern}}, \bibinfo {author} {\bibfnamefont {L.~J.}\ \bibnamefont {Dixon}},\ and\ \bibinfo {author} {\bibfnamefont {D.~A.}\ \bibnamefont {Kosower}},\ }\bibfield  {title} {\bibinfo {title} {{Two-loop $g \to gg$ splitting amplitudes in QCD}},\ }\href {https://doi.org/10.1088/1126-6708/2004/08/012} {\bibfield  {journal} {\bibinfo  {journal} {JHEP}\ }\textbf {\bibinfo {volume} {08}},\ \bibinfo {pages} {012}},\ \Eprint {https://arxiv.org/abs/hep-ph/0404293} {arXiv:hep-ph/0404293} \BibitemShut {NoStop}%
\bibitem [{\citenamefont {Feige}\ and\ \citenamefont {Schwartz}(2014)}]{Feige:2014wja}%
  \BibitemOpen
  \bibfield  {author} {\bibinfo {author} {\bibfnamefont {I.}~\bibnamefont {Feige}}\ and\ \bibinfo {author} {\bibfnamefont {M.~D.}\ \bibnamefont {Schwartz}},\ }\bibfield  {title} {\bibinfo {title} {{Hard-Soft-Collinear Factorization to All Orders}},\ }\href {https://doi.org/10.1103/PhysRevD.90.105020} {\bibfield  {journal} {\bibinfo  {journal} {Phys. Rev. D}\ }\textbf {\bibinfo {volume} {90}},\ \bibinfo {pages} {105020} (\bibinfo {year} {2014})},\ \Eprint {https://arxiv.org/abs/1403.6472} {arXiv:1403.6472 [hep-ph]} \BibitemShut {NoStop}%
\bibitem [{\citenamefont {Guan}\ \emph {et~al.}(2025)\citenamefont {Guan}, \citenamefont {Herzog}, \citenamefont {Ma}, \citenamefont {Mistlberger},\ and\ \citenamefont {Suresh}}]{Guan:2024hlf}%
  \BibitemOpen
  \bibfield  {author} {\bibinfo {author} {\bibfnamefont {X.}~\bibnamefont {Guan}}, \bibinfo {author} {\bibfnamefont {F.}~\bibnamefont {Herzog}}, \bibinfo {author} {\bibfnamefont {Y.}~\bibnamefont {Ma}}, \bibinfo {author} {\bibfnamefont {B.}~\bibnamefont {Mistlberger}},\ and\ \bibinfo {author} {\bibfnamefont {A.}~\bibnamefont {Suresh}},\ }\bibfield  {title} {\bibinfo {title} {{Splitting amplitudes at N$^{3}$LO in QCD}},\ }\href {https://doi.org/10.1007/JHEP01(2025)090} {\bibfield  {journal} {\bibinfo  {journal} {JHEP}\ }\textbf {\bibinfo {volume} {01}},\ \bibinfo {pages} {090}},\ \Eprint {https://arxiv.org/abs/2408.03019} {arXiv:2408.03019 [hep-ph]} \BibitemShut {NoStop}%
\bibitem [{\citenamefont {Duhr}\ \emph {et~al.}(2026)\citenamefont {Duhr}, \citenamefont {Gardi}, \citenamefont {Jaskiewicz}, \citenamefont {L{\"u}bken},\ and\ \citenamefont {Vernazza}}]{Duhr:2025cye}%
  \BibitemOpen
  \bibfield  {author} {\bibinfo {author} {\bibfnamefont {C.}~\bibnamefont {Duhr}}, \bibinfo {author} {\bibfnamefont {E.}~\bibnamefont {Gardi}}, \bibinfo {author} {\bibfnamefont {S.}~\bibnamefont {Jaskiewicz}}, \bibinfo {author} {\bibfnamefont {J.}~\bibnamefont {L{\"u}bken}},\ and\ \bibinfo {author} {\bibfnamefont {L.}~\bibnamefont {Vernazza}},\ }\bibfield  {title} {\bibinfo {title} {{Infrared singularities and the collinear limits of multi-leg scattering amplitudes}},\ }\href {https://doi.org/10.1007/JHEP02(2026)173} {\bibfield  {journal} {\bibinfo  {journal} {JHEP}\ }\textbf {\bibinfo {volume} {02}},\ \bibinfo {pages} {173}},\ \Eprint {https://arxiv.org/abs/2507.21854} {arXiv:2507.21854 [hep-ph]} \BibitemShut {NoStop}%
\bibitem [{\citenamefont {Catani}\ \emph {et~al.}(2012)\citenamefont {Catani}, \citenamefont {de~Florian},\ and\ \citenamefont {Rodrigo}}]{Catani:2011st}%
  \BibitemOpen
  \bibfield  {author} {\bibinfo {author} {\bibfnamefont {S.}~\bibnamefont {Catani}}, \bibinfo {author} {\bibfnamefont {D.}~\bibnamefont {de~Florian}},\ and\ \bibinfo {author} {\bibfnamefont {G.}~\bibnamefont {Rodrigo}},\ }\bibfield  {title} {\bibinfo {title} {{Space-like (versus time-like) collinear limits in QCD: Is factorization violated?}},\ }\href {https://doi.org/10.1007/JHEP07(2012)026} {\bibfield  {journal} {\bibinfo  {journal} {JHEP}\ }\textbf {\bibinfo {volume} {07}},\ \bibinfo {pages} {026}},\ \Eprint {https://arxiv.org/abs/1112.4405} {arXiv:1112.4405 [hep-ph]} \BibitemShut {NoStop}%
\bibitem [{\citenamefont {Henn}\ \emph {et~al.}(2025)\citenamefont {Henn}, \citenamefont {Ma}, \citenamefont {Xu}, \citenamefont {Yan}, \citenamefont {Zhang},\ and\ \citenamefont {Zhu}}]{Henn:2024qjq}%
  \BibitemOpen
  \bibfield  {author} {\bibinfo {author} {\bibfnamefont {J.}~\bibnamefont {Henn}}, \bibinfo {author} {\bibfnamefont {R.}~\bibnamefont {Ma}}, \bibinfo {author} {\bibfnamefont {Y.}~\bibnamefont {Xu}}, \bibinfo {author} {\bibfnamefont {K.}~\bibnamefont {Yan}}, \bibinfo {author} {\bibfnamefont {Y.}~\bibnamefont {Zhang}},\ and\ \bibinfo {author} {\bibfnamefont {H.~X.}\ \bibnamefont {Zhu}},\ }\bibfield  {title} {\bibinfo {title} {{Two-loop spacelike splitting amplitude for $\mathcal{N}=4$ Super-Yang-Mills theory}},\ }\href {https://doi.org/10.1103/5qrc-4vq1} {\bibfield  {journal} {\bibinfo  {journal} {Phys. Rev. D}\ }\textbf {\bibinfo {volume} {112}},\ \bibinfo {pages} {076003} (\bibinfo {year} {2025})},\ \Eprint {https://arxiv.org/abs/2406.14604} {arXiv:2406.14604 [hep-th]} \BibitemShut {NoStop}%
\bibitem [{\citenamefont {Duhr}\ \emph {et~al.}(2025)\citenamefont {Duhr}, \citenamefont {Venkata},\ and\ \citenamefont {Zhang}}]{Duhr:2025lyg}%
  \BibitemOpen
  \bibfield  {author} {\bibinfo {author} {\bibfnamefont {C.}~\bibnamefont {Duhr}}, \bibinfo {author} {\bibfnamefont {A.}~\bibnamefont {Venkata}},\ and\ \bibinfo {author} {\bibfnamefont {C.}~\bibnamefont {Zhang}},\ }\bibfield  {title} {\bibinfo {title} {{Double Spacelike Collinear Limits from Multi-Regge Kinematics}},\ }\href {https://doi.org/10.1103/3bxb-zfwc} {\bibfield  {journal} {\bibinfo  {journal} {Phys. Rev. Lett.}\ }\textbf {\bibinfo {volume} {135}},\ \bibinfo {pages} {241601} (\bibinfo {year} {2025})},\ \Eprint {https://arxiv.org/abs/2507.05355} {arXiv:2507.05355 [hep-th]} \BibitemShut {NoStop}%
\bibitem [{\citenamefont {Buccioni}\ \emph {et~al.}(2026)\citenamefont {Buccioni}, \citenamefont {Fang},\ and\ \citenamefont {Yan}}]{Buccioni:2026mfg}%
  \BibitemOpen
  \bibfield  {author} {\bibinfo {author} {\bibfnamefont {F.}~\bibnamefont {Buccioni}}, \bibinfo {author} {\bibfnamefont {H.}~\bibnamefont {Fang}},\ and\ \bibinfo {author} {\bibfnamefont {K.}~\bibnamefont {Yan}},\ }\href@noop {} {\bibinfo {title} {{Two-Loop Spacelike Splitting Amplitudes in Full-Color QCD}}} (\bibinfo {year} {2026}),\ \Eprint {https://arxiv.org/abs/2603.27123} {arXiv:2603.27123 [hep-ph]} \BibitemShut {NoStop}%
\bibitem [{\citenamefont {Becher}\ \emph {et~al.}(2026{\natexlab{a}})\citenamefont {Becher}, \citenamefont {Hager}, \citenamefont {Neubert},\ and\ \citenamefont {Schwienbacher}}]{splittings}%
  \BibitemOpen
  \bibfield  {author} {\bibinfo {author} {\bibfnamefont {T.}~\bibnamefont {Becher}}, \bibinfo {author} {\bibfnamefont {P.}~\bibnamefont {Hager}}, \bibinfo {author} {\bibfnamefont {M.}~\bibnamefont {Neubert}},\ and\ \bibinfo {author} {\bibfnamefont {D.}~\bibnamefont {Schwienbacher}},\ }\href@noop {} {\bibinfo {title} {{in preparation}}} (\bibinfo {year} {2026}{\natexlab{a}})\BibitemShut {NoStop}%
\bibitem [{\citenamefont {Forshaw}\ \emph {et~al.}(2006)\citenamefont {Forshaw}, \citenamefont {Kyrieleis},\ and\ \citenamefont {Seymour}}]{Forshaw:2006fk}%
  \BibitemOpen
  \bibfield  {author} {\bibinfo {author} {\bibfnamefont {J.~R.}\ \bibnamefont {Forshaw}}, \bibinfo {author} {\bibfnamefont {A.}~\bibnamefont {Kyrieleis}},\ and\ \bibinfo {author} {\bibfnamefont {M.~H.}\ \bibnamefont {Seymour}},\ }\bibfield  {title} {\bibinfo {title} {{Super-leading logarithms in non-global observables in QCD}},\ }\href {https://doi.org/10.1088/1126-6708/2006/08/059} {\bibfield  {journal} {\bibinfo  {journal} {JHEP}\ }\textbf {\bibinfo {volume} {08}},\ \bibinfo {pages} {059}},\ \Eprint {https://arxiv.org/abs/hep-ph/0604094} {arXiv:hep-ph/0604094} \BibitemShut {NoStop}%
\bibitem [{\citenamefont {Forshaw}\ \emph {et~al.}(2008)\citenamefont {Forshaw}, \citenamefont {Kyrieleis},\ and\ \citenamefont {Seymour}}]{Forshaw:2008cq}%
  \BibitemOpen
  \bibfield  {author} {\bibinfo {author} {\bibfnamefont {J.~R.}\ \bibnamefont {Forshaw}}, \bibinfo {author} {\bibfnamefont {A.}~\bibnamefont {Kyrieleis}},\ and\ \bibinfo {author} {\bibfnamefont {M.~H.}\ \bibnamefont {Seymour}},\ }\bibfield  {title} {\bibinfo {title} {{Super-leading logarithms in non-global observables in QCD: Colour basis independent calculation}},\ }\href {https://doi.org/10.1088/1126-6708/2008/09/128} {\bibfield  {journal} {\bibinfo  {journal} {JHEP}\ }\textbf {\bibinfo {volume} {09}},\ \bibinfo {pages} {128}},\ \Eprint {https://arxiv.org/abs/0808.1269} {arXiv:0808.1269 [hep-ph]} \BibitemShut {NoStop}%
\bibitem [{\citenamefont {Dasgupta}\ and\ \citenamefont {Salam}(2001)}]{Dasgupta:2001sh}%
  \BibitemOpen
  \bibfield  {author} {\bibinfo {author} {\bibfnamefont {M.}~\bibnamefont {Dasgupta}}\ and\ \bibinfo {author} {\bibfnamefont {G.~P.}\ \bibnamefont {Salam}},\ }\bibfield  {title} {\bibinfo {title} {{Resummation of nonglobal QCD observables}},\ }\href {https://doi.org/10.1016/S0370-2693(01)00725-0} {\bibfield  {journal} {\bibinfo  {journal} {Phys. Lett. B}\ }\textbf {\bibinfo {volume} {512}},\ \bibinfo {pages} {323} (\bibinfo {year} {2001})},\ \Eprint {https://arxiv.org/abs/hep-ph/0104277} {arXiv:hep-ph/0104277} \BibitemShut {NoStop}%
\bibitem [{\citenamefont {Becher}\ \emph {et~al.}(2021)\citenamefont {Becher}, \citenamefont {Neubert},\ and\ \citenamefont {Shao}}]{Becher:2021zkk}%
  \BibitemOpen
  \bibfield  {author} {\bibinfo {author} {\bibfnamefont {T.}~\bibnamefont {Becher}}, \bibinfo {author} {\bibfnamefont {M.}~\bibnamefont {Neubert}},\ and\ \bibinfo {author} {\bibfnamefont {D.~Y.}\ \bibnamefont {Shao}},\ }\bibfield  {title} {\bibinfo {title} {{Resummation of Super-Leading Logarithms}},\ }\href {https://doi.org/10.1103/PhysRevLett.127.212002} {\bibfield  {journal} {\bibinfo  {journal} {Phys. Rev. Lett.}\ }\textbf {\bibinfo {volume} {127}},\ \bibinfo {pages} {212002} (\bibinfo {year} {2021})},\ \Eprint {https://arxiv.org/abs/2107.01212} {arXiv:2107.01212 [hep-ph]} \BibitemShut {NoStop}%
\bibitem [{\citenamefont {Becher}\ \emph {et~al.}(2023)\citenamefont {Becher}, \citenamefont {Neubert}, \citenamefont {Shao},\ and\ \citenamefont {Stillger}}]{Becher:2023mtx}%
  \BibitemOpen
  \bibfield  {author} {\bibinfo {author} {\bibfnamefont {T.}~\bibnamefont {Becher}}, \bibinfo {author} {\bibfnamefont {M.}~\bibnamefont {Neubert}}, \bibinfo {author} {\bibfnamefont {D.~Y.}\ \bibnamefont {Shao}},\ and\ \bibinfo {author} {\bibfnamefont {M.}~\bibnamefont {Stillger}},\ }\bibfield  {title} {\bibinfo {title} {{Factorization of non-global LHC observables and resummation of super-leading logarithms}},\ }\href {https://doi.org/10.1007/JHEP12(2023)116} {\bibfield  {journal} {\bibinfo  {journal} {JHEP}\ }\textbf {\bibinfo {volume} {12}},\ \bibinfo {pages} {116}},\ \Eprint {https://arxiv.org/abs/2307.06359} {arXiv:2307.06359 [hep-ph]} \BibitemShut {NoStop}%
\bibitem [{\citenamefont {B{\"o}er}\ \emph {et~al.}(2023)\citenamefont {B{\"o}er}, \citenamefont {Neubert},\ and\ \citenamefont {Stillger}}]{Boer:2023jsy}%
  \BibitemOpen
  \bibfield  {author} {\bibinfo {author} {\bibfnamefont {P.}~\bibnamefont {B{\"o}er}}, \bibinfo {author} {\bibfnamefont {M.}~\bibnamefont {Neubert}},\ and\ \bibinfo {author} {\bibfnamefont {M.}~\bibnamefont {Stillger}},\ }\bibfield  {title} {\bibinfo {title} {{Glauber phases in non-global LHC observables: resummation for quark-initiated processes}},\ }\href {https://doi.org/10.1007/JHEP10(2023)075} {\bibfield  {journal} {\bibinfo  {journal} {JHEP}\ }\textbf {\bibinfo {volume} {10}},\ \bibinfo {pages} {075}},\ \Eprint {https://arxiv.org/abs/2307.11089} {arXiv:2307.11089 [hep-ph]} \BibitemShut {NoStop}%
\bibitem [{\citenamefont {B{\"o}er}\ \emph {et~al.}(2024{\natexlab{a}})\citenamefont {B{\"o}er}, \citenamefont {Hager}, \citenamefont {Neubert}, \citenamefont {Stillger},\ and\ \citenamefont {Xu}}]{Boer:2023ljq}%
  \BibitemOpen
  \bibfield  {author} {\bibinfo {author} {\bibfnamefont {P.}~\bibnamefont {B{\"o}er}}, \bibinfo {author} {\bibfnamefont {P.}~\bibnamefont {Hager}}, \bibinfo {author} {\bibfnamefont {M.}~\bibnamefont {Neubert}}, \bibinfo {author} {\bibfnamefont {M.}~\bibnamefont {Stillger}},\ and\ \bibinfo {author} {\bibfnamefont {X.}~\bibnamefont {Xu}},\ }\bibfield  {title} {\bibinfo {title} {{Glauber phases in non-global LHC observables: resummation for gluon-initiated processes}},\ }\href {https://doi.org/10.1007/JHEP02(2024)109} {\bibfield  {journal} {\bibinfo  {journal} {JHEP}\ }\textbf {\bibinfo {volume} {02}},\ \bibinfo {pages} {109}},\ \Eprint {https://arxiv.org/abs/2311.18811} {arXiv:2311.18811 [hep-ph]} \BibitemShut {NoStop}%
\bibitem [{\citenamefont {B{\"o}er}\ \emph {et~al.}(2024{\natexlab{b}})\citenamefont {B{\"o}er}, \citenamefont {Hager}, \citenamefont {Neubert}, \citenamefont {Stillger},\ and\ \citenamefont {Xu}}]{Boer:2024xzy}%
  \BibitemOpen
  \bibfield  {author} {\bibinfo {author} {\bibfnamefont {P.}~\bibnamefont {B{\"o}er}}, \bibinfo {author} {\bibfnamefont {P.}~\bibnamefont {Hager}}, \bibinfo {author} {\bibfnamefont {M.}~\bibnamefont {Neubert}}, \bibinfo {author} {\bibfnamefont {M.}~\bibnamefont {Stillger}},\ and\ \bibinfo {author} {\bibfnamefont {X.}~\bibnamefont {Xu}},\ }\bibfield  {title} {\bibinfo {title} {{Resummation of Glauber phases in non-global LHC observables for large $N_{c}$}},\ }\href {https://doi.org/10.1007/JHEP08(2024)036} {\bibfield  {journal} {\bibinfo  {journal} {JHEP}\ }\textbf {\bibinfo {volume} {08}},\ \bibinfo {pages} {036}},\ \Eprint {https://arxiv.org/abs/2407.01691} {arXiv:2407.01691 [hep-ph]} \BibitemShut {NoStop}%
\bibitem [{\citenamefont {Banerjee}\ \emph {et~al.}(2026)\citenamefont {Banerjee}, \citenamefont {Gr{\"u}nhofer}, \citenamefont {K{\"o}nig}, \citenamefont {Li}, \citenamefont {Neubert},\ and\ \citenamefont {Scholze}}]{Banerjee:2025kkq}%
  \BibitemOpen
  \bibfield  {author} {\bibinfo {author} {\bibfnamefont {U.}~\bibnamefont {Banerjee}}, \bibinfo {author} {\bibfnamefont {R.}~\bibnamefont {Gr{\"u}nhofer}}, \bibinfo {author} {\bibfnamefont {M.}~\bibnamefont {K{\"o}nig}}, \bibinfo {author} {\bibfnamefont {Y.}~\bibnamefont {Li}}, \bibinfo {author} {\bibfnamefont {M.}~\bibnamefont {Neubert}},\ and\ \bibinfo {author} {\bibfnamefont {J.}~\bibnamefont {Scholze}},\ }\bibfield  {title} {\bibinfo {title} {{Super-leading logarithms in top-quark pair production at hadron colliders}},\ }\href {https://doi.org/10.1007/JHEP02(2026)092} {\bibfield  {journal} {\bibinfo  {journal} {JHEP}\ }\textbf {\bibinfo {volume} {02}},\ \bibinfo {pages} {092}},\ \Eprint {https://arxiv.org/abs/2510.24848} {arXiv:2510.24848 [hep-ph]} \BibitemShut {NoStop}%
\bibitem [{\citenamefont {Bauer}\ \emph {et~al.}(2002{\natexlab{a}})\citenamefont {Bauer}, \citenamefont {Pirjol},\ and\ \citenamefont {Stewart}}]{Bauer:2001yt}%
  \BibitemOpen
  \bibfield  {author} {\bibinfo {author} {\bibfnamefont {C.~W.}\ \bibnamefont {Bauer}}, \bibinfo {author} {\bibfnamefont {D.}~\bibnamefont {Pirjol}},\ and\ \bibinfo {author} {\bibfnamefont {I.~W.}\ \bibnamefont {Stewart}},\ }\bibfield  {title} {\bibinfo {title} {{Soft collinear factorization in effective field theory}},\ }\href {https://doi.org/10.1103/PhysRevD.65.054022} {\bibfield  {journal} {\bibinfo  {journal} {Phys. Rev. D}\ }\textbf {\bibinfo {volume} {65}},\ \bibinfo {pages} {054022} (\bibinfo {year} {2002}{\natexlab{a}})},\ \Eprint {https://arxiv.org/abs/hep-ph/0109045} {arXiv:hep-ph/0109045} \BibitemShut {NoStop}%
\bibitem [{\citenamefont {Bauer}\ \emph {et~al.}(2002{\natexlab{b}})\citenamefont {Bauer}, \citenamefont {Fleming}, \citenamefont {Pirjol}, \citenamefont {Rothstein},\ and\ \citenamefont {Stewart}}]{Bauer:2002nz}%
  \BibitemOpen
  \bibfield  {author} {\bibinfo {author} {\bibfnamefont {C.~W.}\ \bibnamefont {Bauer}}, \bibinfo {author} {\bibfnamefont {S.}~\bibnamefont {Fleming}}, \bibinfo {author} {\bibfnamefont {D.}~\bibnamefont {Pirjol}}, \bibinfo {author} {\bibfnamefont {I.~Z.}\ \bibnamefont {Rothstein}},\ and\ \bibinfo {author} {\bibfnamefont {I.~W.}\ \bibnamefont {Stewart}},\ }\bibfield  {title} {\bibinfo {title} {{Hard scattering factorization from effective field theory}},\ }\href {https://doi.org/10.1103/PhysRevD.66.014017} {\bibfield  {journal} {\bibinfo  {journal} {Phys. Rev. D}\ }\textbf {\bibinfo {volume} {66}},\ \bibinfo {pages} {014017} (\bibinfo {year} {2002}{\natexlab{b}})},\ \Eprint {https://arxiv.org/abs/hep-ph/0202088} {arXiv:hep-ph/0202088} \BibitemShut {NoStop}%
\bibitem [{\citenamefont {Beneke}\ \emph {et~al.}(2002)\citenamefont {Beneke}, \citenamefont {Chapovsky}, \citenamefont {Diehl},\ and\ \citenamefont {Feldmann}}]{Beneke:2002ph}%
  \BibitemOpen
  \bibfield  {author} {\bibinfo {author} {\bibfnamefont {M.}~\bibnamefont {Beneke}}, \bibinfo {author} {\bibfnamefont {A.~P.}\ \bibnamefont {Chapovsky}}, \bibinfo {author} {\bibfnamefont {M.}~\bibnamefont {Diehl}},\ and\ \bibinfo {author} {\bibfnamefont {T.}~\bibnamefont {Feldmann}},\ }\bibfield  {title} {\bibinfo {title} {{Soft collinear effective theory and heavy to light currents beyond leading power}},\ }\href {https://doi.org/10.1016/S0550-3213(02)00687-9} {\bibfield  {journal} {\bibinfo  {journal} {Nucl. Phys. B}\ }\textbf {\bibinfo {volume} {643}},\ \bibinfo {pages} {431} (\bibinfo {year} {2002})},\ \Eprint {https://arxiv.org/abs/hep-ph/0206152} {arXiv:hep-ph/0206152} \BibitemShut {NoStop}%
\bibitem [{\citenamefont {Beneke}\ and\ \citenamefont {Feldmann}(2003)}]{Beneke:2002ni}%
  \BibitemOpen
  \bibfield  {author} {\bibinfo {author} {\bibfnamefont {M.}~\bibnamefont {Beneke}}\ and\ \bibinfo {author} {\bibfnamefont {T.}~\bibnamefont {Feldmann}},\ }\bibfield  {title} {\bibinfo {title} {{Multipole expanded soft collinear effective theory with non-Abelian gauge symmetry}},\ }\href {https://doi.org/10.1016/S0370-2693(02)03204-5} {\bibfield  {journal} {\bibinfo  {journal} {Phys. Lett. B}\ }\textbf {\bibinfo {volume} {553}},\ \bibinfo {pages} {267} (\bibinfo {year} {2003})},\ \Eprint {https://arxiv.org/abs/hep-ph/0211358} {arXiv:hep-ph/0211358} \BibitemShut {NoStop}%
\bibitem [{\citenamefont {Becher}\ \emph {et~al.}(2025{\natexlab{a}})\citenamefont {Becher}, \citenamefont {Hager}, \citenamefont {Martinelli}, \citenamefont {Neubert}, \citenamefont {Schwienbacher},\ and\ \citenamefont {Stillger}}]{Becher:2024nqc}%
  \BibitemOpen
  \bibfield  {author} {\bibinfo {author} {\bibfnamefont {T.}~\bibnamefont {Becher}}, \bibinfo {author} {\bibfnamefont {P.}~\bibnamefont {Hager}}, \bibinfo {author} {\bibfnamefont {G.}~\bibnamefont {Martinelli}}, \bibinfo {author} {\bibfnamefont {M.}~\bibnamefont {Neubert}}, \bibinfo {author} {\bibfnamefont {D.}~\bibnamefont {Schwienbacher}},\ and\ \bibinfo {author} {\bibfnamefont {M.}~\bibnamefont {Stillger}},\ }\bibfield  {title} {\bibinfo {title} {{Super-leading logarithms in $pp \to 2$ jets}},\ }\href {https://doi.org/10.1007/JHEP01(2025)171} {\bibfield  {journal} {\bibinfo  {journal} {JHEP}\ }\textbf {\bibinfo {volume} {01}},\ \bibinfo {pages} {171}},\ \Eprint {https://arxiv.org/abs/2411.12742} {arXiv:2411.12742 [hep-ph]} \BibitemShut {NoStop}%
\bibitem [{\citenamefont {Forshaw}\ \emph {et~al.}(2026)\citenamefont {Forshaw}, \citenamefont {Pl{\"a}tzer},\ and\ \citenamefont {Gonz{\'a}lez}}]{Forshaw:2025bmo}%
  \BibitemOpen
  \bibfield  {author} {\bibinfo {author} {\bibfnamefont {J.~R.}\ \bibnamefont {Forshaw}}, \bibinfo {author} {\bibfnamefont {S.}~\bibnamefont {Pl{\"a}tzer}},\ and\ \bibinfo {author} {\bibfnamefont {F.~T.}\ \bibnamefont {Gonz{\'a}lez}},\ }\bibfield  {title} {\bibinfo {title} {{Exact colour evolution for jet observables}},\ }\href {https://doi.org/10.1007/JHEP02(2026)174} {\bibfield  {journal} {\bibinfo  {journal} {JHEP}\ }\textbf {\bibinfo {volume} {02}},\ \bibinfo {pages} {174}},\ \Eprint {https://arxiv.org/abs/2502.12133} {arXiv:2502.12133 [hep-ph]} \BibitemShut {NoStop}%
\bibitem [{\citenamefont {Forshaw}\ \emph {et~al.}(2025)\citenamefont {Forshaw}, \citenamefont {Pl{\"a}tzer},\ and\ \citenamefont {Gonz{\'a}lez}}]{Forshaw:2025fif}%
  \BibitemOpen
  \bibfield  {author} {\bibinfo {author} {\bibfnamefont {J.~R.}\ \bibnamefont {Forshaw}}, \bibinfo {author} {\bibfnamefont {S.}~\bibnamefont {Pl{\"a}tzer}},\ and\ \bibinfo {author} {\bibfnamefont {F.~T.}\ \bibnamefont {Gonz{\'a}lez}},\ }\bibfield  {title} {\bibinfo {title} {{Fully differential soft gluon evolution at the amplitude level}},\ }\href {https://doi.org/10.1007/JHEP09(2025)207} {\bibfield  {journal} {\bibinfo  {journal} {JHEP}\ }\textbf {\bibinfo {volume} {09}},\ \bibinfo {pages} {207}},\ \Eprint {https://arxiv.org/abs/2505.13183} {arXiv:2505.13183 [hep-ph]} \BibitemShut {NoStop}%
\bibitem [{\citenamefont {Dasgupta}\ \emph {et~al.}(2026{\natexlab{a}})\citenamefont {Dasgupta}, \citenamefont {Fraley}, \citenamefont {Monni},\ and\ \citenamefont {Nabeebaccus}}]{Dasgupta:2025cgl}%
  \BibitemOpen
  \bibfield  {author} {\bibinfo {author} {\bibfnamefont {M.}~\bibnamefont {Dasgupta}}, \bibinfo {author} {\bibfnamefont {A.}~\bibnamefont {Fraley}}, \bibinfo {author} {\bibfnamefont {P.~F.}\ \bibnamefont {Monni}},\ and\ \bibinfo {author} {\bibfnamefont {S.}~\bibnamefont {Nabeebaccus}},\ }\bibfield  {title} {\bibinfo {title} {{Uncharted Logarithmic Structures in QCD Transverse-Energy Flow}},\ }\href {https://doi.org/10.1103/6bv2-ybsg} {\bibfield  {journal} {\bibinfo  {journal} {Phys. Rev. Lett.}\ }\textbf {\bibinfo {volume} {137}},\ \bibinfo {pages} {111906} (\bibinfo {year} {2026}{\natexlab{a}})},\ \Eprint {https://arxiv.org/abs/2511.14681} {arXiv:2511.14681 [hep-ph]} \BibitemShut {NoStop}%
\bibitem [{\citenamefont {Becher}\ \emph {et~al.}(2026{\natexlab{b}})\citenamefont {Becher}, \citenamefont {Hager}, \citenamefont {Neubert}, \citenamefont {Schalch}, \citenamefont {Scholze},\ and\ \citenamefont {Schwienbacher}}]{Becher:2026uih}%
  \BibitemOpen
  \bibfield  {author} {\bibinfo {author} {\bibfnamefont {T.}~\bibnamefont {Becher}}, \bibinfo {author} {\bibfnamefont {P.}~\bibnamefont {Hager}}, \bibinfo {author} {\bibfnamefont {M.}~\bibnamefont {Neubert}}, \bibinfo {author} {\bibfnamefont {N.}~\bibnamefont {Schalch}}, \bibinfo {author} {\bibfnamefont {J.}~\bibnamefont {Scholze}},\ and\ \bibinfo {author} {\bibfnamefont {D.}~\bibnamefont {Schwienbacher}},\ }\href@noop {} {\bibinfo {title} {{Non-Global and Super-Leading Logarithms in Higgs Production through Vector Boson Fusion}}} (\bibinfo {year} {2026}{\natexlab{b}}),\ \Eprint {https://arxiv.org/abs/2611.03407} {arXiv:2611.03407 [hep-ph]} \BibitemShut {NoStop}%
\bibitem [{\citenamefont {Schalch}(2026)}]{Schalch:2026xxk}%
  \BibitemOpen
  \bibfield  {author} {\bibinfo {author} {\bibfnamefont {N.}~\bibnamefont {Schalch}},\ }\href@noop {} {\bibinfo {title} {{A First Glimpse at Super-Leading Logarithms in Monte Carlo Evolution}}} (\bibinfo {year} {2026}),\ \Eprint {https://arxiv.org/abs/2609.34984} {arXiv:2609.34984 [hep-ph]} \BibitemShut {NoStop}%
\bibitem [{\citenamefont {Dasgupta}\ \emph {et~al.}(2026{\natexlab{b}})\citenamefont {Dasgupta}, \citenamefont {Fraley}, \citenamefont {Monni},\ and\ \citenamefont {Nabeebaccus}}]{Dasgupta:2026mlm}%
  \BibitemOpen
  \bibfield  {author} {\bibinfo {author} {\bibfnamefont {M.}~\bibnamefont {Dasgupta}}, \bibinfo {author} {\bibfnamefont {A.}~\bibnamefont {Fraley}}, \bibinfo {author} {\bibfnamefont {P.~F.}\ \bibnamefont {Monni}},\ and\ \bibinfo {author} {\bibfnamefont {S.}~\bibnamefont {Nabeebaccus}},\ }\href@noop {} {\bibinfo {title} {{A new angle on coherence-violating logarithms in color-singlet processes}}} (\bibinfo {year} {2026}{\natexlab{b}}),\ \Eprint {https://arxiv.org/abs/2611.21763} {arXiv:2611.21763 [hep-ph]} \BibitemShut {NoStop}%
\bibitem [{\citenamefont {Banfi}\ \emph {et~al.}(2010)\citenamefont {Banfi}, \citenamefont {Salam},\ and\ \citenamefont {Zanderighi}}]{Banfi:2010xy}%
  \BibitemOpen
  \bibfield  {author} {\bibinfo {author} {\bibfnamefont {A.}~\bibnamefont {Banfi}}, \bibinfo {author} {\bibfnamefont {G.~P.}\ \bibnamefont {Salam}},\ and\ \bibinfo {author} {\bibfnamefont {G.}~\bibnamefont {Zanderighi}},\ }\bibfield  {title} {\bibinfo {title} {{Phenomenology of event shapes at hadron colliders}},\ }\href {https://doi.org/10.1007/JHEP06(2010)038} {\bibfield  {journal} {\bibinfo  {journal} {JHEP}\ }\textbf {\bibinfo {volume} {06}},\ \bibinfo {pages} {038}},\ \Eprint {https://arxiv.org/abs/1001.4082} {arXiv:1001.4082 [hep-ph]} \BibitemShut {NoStop}%
\bibitem [{\citenamefont {Forshaw}\ and\ \citenamefont {Holguin}(2021)}]{Forshaw:2021fxs}%
  \BibitemOpen
  \bibfield  {author} {\bibinfo {author} {\bibfnamefont {J.~R.}\ \bibnamefont {Forshaw}}\ and\ \bibinfo {author} {\bibfnamefont {J.}~\bibnamefont {Holguin}},\ }\bibfield  {title} {\bibinfo {title} {{Coulomb gluons will generally destroy coherence}},\ }\href {https://doi.org/10.1007/JHEP12(2021)084} {\bibfield  {journal} {\bibinfo  {journal} {JHEP}\ }\textbf {\bibinfo {volume} {12}},\ \bibinfo {pages} {084}},\ \bibinfo {note} {[Erratum: JHEP 04, 097 (2024)]},\ \Eprint {https://arxiv.org/abs/2109.03665} {arXiv:2109.03665 [hep-ph]} \BibitemShut {NoStop}%
\bibitem [{\citenamefont {Beneke}\ and\ \citenamefont {Smirnov}(1998)}]{Beneke:1997zp}%
  \BibitemOpen
  \bibfield  {author} {\bibinfo {author} {\bibfnamefont {M.}~\bibnamefont {Beneke}}\ and\ \bibinfo {author} {\bibfnamefont {V.~A.}\ \bibnamefont {Smirnov}},\ }\bibfield  {title} {\bibinfo {title} {{Asymptotic expansion of Feynman integrals near threshold}},\ }\href {https://doi.org/10.1016/S0550-3213(98)00138-2} {\bibfield  {journal} {\bibinfo  {journal} {Nucl. Phys. B}\ }\textbf {\bibinfo {volume} {522}},\ \bibinfo {pages} {321} (\bibinfo {year} {1998})},\ \Eprint {https://arxiv.org/abs/hep-ph/9711391} {arXiv:hep-ph/9711391} \BibitemShut {NoStop}%
\bibitem [{\citenamefont {Pak}\ and\ \citenamefont {Smirnov}(2011)}]{Pak:2010pt}%
  \BibitemOpen
  \bibfield  {author} {\bibinfo {author} {\bibfnamefont {A.}~\bibnamefont {Pak}}\ and\ \bibinfo {author} {\bibfnamefont {A.}~\bibnamefont {Smirnov}},\ }\bibfield  {title} {\bibinfo {title} {{Geometric approach to asymptotic expansion of Feynman integrals}},\ }\href {https://doi.org/10.1140/epjc/s10052-011-1626-1} {\bibfield  {journal} {\bibinfo  {journal} {Eur. Phys. J. C}\ }\textbf {\bibinfo {volume} {71}},\ \bibinfo {pages} {1626} (\bibinfo {year} {2011})},\ \Eprint {https://arxiv.org/abs/1011.4863} {arXiv:1011.4863 [hep-ph]} \BibitemShut {NoStop}%
\bibitem [{\citenamefont {Jantzen}(2011)}]{Jantzen:2011nz}%
  \BibitemOpen
  \bibfield  {author} {\bibinfo {author} {\bibfnamefont {B.}~\bibnamefont {Jantzen}},\ }\bibfield  {title} {\bibinfo {title} {{Foundation and generalization of the expansion by regions}},\ }\href {https://doi.org/10.1007/JHEP12(2011)076} {\bibfield  {journal} {\bibinfo  {journal} {JHEP}\ }\textbf {\bibinfo {volume} {12}},\ \bibinfo {pages} {076}},\ \Eprint {https://arxiv.org/abs/1111.2589} {arXiv:1111.2589 [hep-ph]} \BibitemShut {NoStop}%
\bibitem [{\citenamefont {Jantzen}\ \emph {et~al.}(2012)\citenamefont {Jantzen}, \citenamefont {Smirnov},\ and\ \citenamefont {Smirnov}}]{Jantzen:2012mw}%
  \BibitemOpen
  \bibfield  {author} {\bibinfo {author} {\bibfnamefont {B.}~\bibnamefont {Jantzen}}, \bibinfo {author} {\bibfnamefont {A.~V.}\ \bibnamefont {Smirnov}},\ and\ \bibinfo {author} {\bibfnamefont {V.~A.}\ \bibnamefont {Smirnov}},\ }\bibfield  {title} {\bibinfo {title} {{Expansion by regions: revealing potential and Glauber regions automatically}},\ }\href {https://doi.org/10.1140/epjc/s10052-012-2139-2} {\bibfield  {journal} {\bibinfo  {journal} {Eur. Phys. J. C}\ }\textbf {\bibinfo {volume} {72}},\ \bibinfo {pages} {2139} (\bibinfo {year} {2012})},\ \Eprint {https://arxiv.org/abs/1206.0546} {arXiv:1206.0546 [hep-ph]} \BibitemShut {NoStop}%
\bibitem [{\citenamefont {Heinrich}\ \emph {et~al.}(2022)\citenamefont {Heinrich}, \citenamefont {Jahn}, \citenamefont {Jones}, \citenamefont {Kerner}, \citenamefont {Langer}, \citenamefont {Magerya}, \citenamefont {P{\"o}ldaru}, \citenamefont {Schlenk},\ and\ \citenamefont {Villa}}]{Heinrich:2021dbf}%
  \BibitemOpen
  \bibfield  {author} {\bibinfo {author} {\bibfnamefont {G.}~\bibnamefont {Heinrich}}, \bibinfo {author} {\bibfnamefont {S.}~\bibnamefont {Jahn}}, \bibinfo {author} {\bibfnamefont {S.~P.}\ \bibnamefont {Jones}}, \bibinfo {author} {\bibfnamefont {M.}~\bibnamefont {Kerner}}, \bibinfo {author} {\bibfnamefont {F.}~\bibnamefont {Langer}}, \bibinfo {author} {\bibfnamefont {V.}~\bibnamefont {Magerya}}, \bibinfo {author} {\bibfnamefont {A.}~\bibnamefont {P{\"o}ldaru}}, \bibinfo {author} {\bibfnamefont {J.}~\bibnamefont {Schlenk}},\ and\ \bibinfo {author} {\bibfnamefont {E.}~\bibnamefont {Villa}},\ }\bibfield  {title} {\bibinfo {title} {{Expansion by regions with pySecDec}},\ }\href {https://doi.org/10.1016/j.cpc.2021.108267} {\bibfield  {journal} {\bibinfo  {journal} {Comput. Phys. Commun.}\ }\textbf {\bibinfo {volume} {273}},\ \bibinfo {pages} {108267} (\bibinfo {year} {2022})},\ \Eprint {https://arxiv.org/abs/2108.10807} {arXiv:2108.10807 [hep-ph]} \BibitemShut {NoStop}%
\bibitem [{\citenamefont {Ma}(2024)}]{Ma:2023hrt}%
  \BibitemOpen
  \bibfield  {author} {\bibinfo {author} {\bibfnamefont {Y.}~\bibnamefont {Ma}},\ }\bibfield  {title} {\bibinfo {title} {{Identifying regions in wide-angle scattering via graph-theoretical approaches}},\ }\href {https://doi.org/10.1007/JHEP09(2024)197} {\bibfield  {journal} {\bibinfo  {journal} {JHEP}\ }\textbf {\bibinfo {volume} {09}},\ \bibinfo {pages} {197}},\ \Eprint {https://arxiv.org/abs/2312.14012} {arXiv:2312.14012 [hep-ph]} \BibitemShut {NoStop}%
\bibitem [{\citenamefont {Gardi}\ \emph {et~al.}(2024)\citenamefont {Gardi}, \citenamefont {Herzog}, \citenamefont {Jones},\ and\ \citenamefont {Ma}}]{Gardi:2024axt}%
  \BibitemOpen
  \bibfield  {author} {\bibinfo {author} {\bibfnamefont {E.}~\bibnamefont {Gardi}}, \bibinfo {author} {\bibfnamefont {F.}~\bibnamefont {Herzog}}, \bibinfo {author} {\bibfnamefont {S.}~\bibnamefont {Jones}},\ and\ \bibinfo {author} {\bibfnamefont {Y.}~\bibnamefont {Ma}},\ }\bibfield  {title} {\bibinfo {title} {{Dissecting polytopes: Landau singularities and asymptotic expansions in $2 \to 2$ scattering}},\ }\href {https://doi.org/10.1007/JHEP08(2024)127} {\bibfield  {journal} {\bibinfo  {journal} {JHEP}\ }\textbf {\bibinfo {volume} {08}},\ \bibinfo {pages} {127}},\ \Eprint {https://arxiv.org/abs/2407.13738} {arXiv:2407.13738 [hep-th]} \BibitemShut {NoStop}%
\bibitem [{\citenamefont {Chen}\ \emph {et~al.}(2026)\citenamefont {Chen}, \citenamefont {Gardi}, \citenamefont {Ma}, \citenamefont {Ma}, \citenamefont {Zhang},\ and\ \citenamefont {Zhu}}]{Chen:2026dnj}%
  \BibitemOpen
  \bibfield  {author} {\bibinfo {author} {\bibfnamefont {W.}~\bibnamefont {Chen}}, \bibinfo {author} {\bibfnamefont {E.}~\bibnamefont {Gardi}}, \bibinfo {author} {\bibfnamefont {R.}~\bibnamefont {Ma}}, \bibinfo {author} {\bibfnamefont {Y.}~\bibnamefont {Ma}}, \bibinfo {author} {\bibfnamefont {Y.}~\bibnamefont {Zhang}},\ and\ \bibinfo {author} {\bibfnamefont {Z.}~\bibnamefont {Zhu}},\ }\href@noop {} {\bibinfo {title} {{Spacelike-Collinear Scattering by the Method of Regions}}} (\bibinfo {year} {2026}),\ \Eprint {https://arxiv.org/abs/2607.15126} {arXiv:2607.15126 [hep-ph]} \BibitemShut {NoStop}%
\bibitem [{\citenamefont {Gardi}(2026)}]{Gardi:2026mcx}%
  \BibitemOpen
  \bibfield  {author} {\bibinfo {author} {\bibfnamefont {E.}~\bibnamefont {Gardi}},\ }\href@noop {} {\bibinfo {title} {{Hidden Region Finder: asymptotic expansions of Feynman integrals in Minkowski space}}} (\bibinfo {year} {2026}),\ \Eprint {https://arxiv.org/abs/2609.29442} {arXiv:2609.29442 [hep-ph]} \BibitemShut {NoStop}%
\bibitem [{\citenamefont {Becher}\ \emph {et~al.}(2026{\natexlab{c}})\citenamefont {Becher}, \citenamefont {Hager}, \citenamefont {Jaskiewicz}, \citenamefont {Neubert},\ and\ \citenamefont {Schwienbacher}}]{Becher:2025igg}%
  \BibitemOpen
  \bibfield  {author} {\bibinfo {author} {\bibfnamefont {T.}~\bibnamefont {Becher}}, \bibinfo {author} {\bibfnamefont {P.}~\bibnamefont {Hager}}, \bibinfo {author} {\bibfnamefont {S.}~\bibnamefont {Jaskiewicz}}, \bibinfo {author} {\bibfnamefont {M.}~\bibnamefont {Neubert}},\ and\ \bibinfo {author} {\bibfnamefont {D.}~\bibnamefont {Schwienbacher}},\ }\bibfield  {title} {\bibinfo {title} {{Low-energy theory of jet processes and PDF factorization}},\ }\href {https://doi.org/10.1007/JHEP01(2026)024} {\bibfield  {journal} {\bibinfo  {journal} {JHEP}\ }\textbf {\bibinfo {volume} {01}},\ \bibinfo {pages} {024}},\ \Eprint {https://arxiv.org/abs/2509.07082} {arXiv:2509.07082 [hep-ph]} \BibitemShut {NoStop}%
\bibitem [{\citenamefont {Becher}\ and\ \citenamefont {Neubert}(2009{\natexlab{a}})}]{Becher:2009cu}%
  \BibitemOpen
  \bibfield  {author} {\bibinfo {author} {\bibfnamefont {T.}~\bibnamefont {Becher}}\ and\ \bibinfo {author} {\bibfnamefont {M.}~\bibnamefont {Neubert}},\ }\bibfield  {title} {\bibinfo {title} {{Infrared singularities of scattering amplitudes in perturbative QCD}},\ }\href {https://doi.org/10.1103/PhysRevLett.102.162001} {\bibfield  {journal} {\bibinfo  {journal} {Phys. Rev. Lett.}\ }\textbf {\bibinfo {volume} {102}},\ \bibinfo {pages} {162001} (\bibinfo {year} {2009}{\natexlab{a}})},\ \bibinfo {note} {[Erratum: Phys.Rev.Lett. 111, 199905 (2013)]},\ \Eprint {https://arxiv.org/abs/0901.0722} {arXiv:0901.0722 [hep-ph]} \BibitemShut {NoStop}%
\bibitem [{\citenamefont {Gardi}\ and\ \citenamefont {Magnea}(2009)}]{Gardi:2009zv}%
  \BibitemOpen
  \bibfield  {author} {\bibinfo {author} {\bibfnamefont {E.}~\bibnamefont {Gardi}}\ and\ \bibinfo {author} {\bibfnamefont {L.}~\bibnamefont {Magnea}},\ }\bibfield  {title} {\bibinfo {title} {{Infrared singularities in QCD amplitudes}},\ }\href {https://doi.org/10.1393/ncc/i2010-10528-x} {\bibfield  {journal} {\bibinfo  {journal} {Nuovo Cim. C}\ }\textbf {\bibinfo {volume} {32N5-6}},\ \bibinfo {pages} {137} (\bibinfo {year} {2009})},\ \Eprint {https://arxiv.org/abs/0908.3273} {arXiv:0908.3273 [hep-ph]} \BibitemShut {NoStop}%
\bibitem [{\citenamefont {Becher}\ and\ \citenamefont {Neubert}(2009{\natexlab{b}})}]{Becher:2009qa}%
  \BibitemOpen
  \bibfield  {author} {\bibinfo {author} {\bibfnamefont {T.}~\bibnamefont {Becher}}\ and\ \bibinfo {author} {\bibfnamefont {M.}~\bibnamefont {Neubert}},\ }\bibfield  {title} {\bibinfo {title} {{On the Structure of Infrared Singularities of Gauge-Theory Amplitudes}},\ }\href {https://doi.org/10.1088/1126-6708/2009/06/081} {\bibfield  {journal} {\bibinfo  {journal} {JHEP}\ }\textbf {\bibinfo {volume} {06}},\ \bibinfo {pages} {081}},\ \bibinfo {note} {[Erratum: JHEP 11, 024 (2013)]},\ \Eprint {https://arxiv.org/abs/0903.1126} {arXiv:0903.1126 [hep-ph]} \BibitemShut {NoStop}%
\bibitem [{Note1()}]{Note1}%
  \BibitemOpen
  \bibinfo {note} {Note that the more intricate structure of the Glauber term discussed in~\cite {Becher:2026kbr} does not impact the general argument made here.}\BibitemShut {Stop}%
\bibitem [{Note2()}]{Note2}%
  \BibitemOpen
  \bibinfo {note} {A pattern reminiscent of SLLs has also been observed at small-$x$ in the four-loop non-singlet quark splitting function~\cite {Davies:2022ofz,Gehrmann:2026qbl,Moch:2026qsw}.}\BibitemShut {Stop}%
\bibitem [{\citenamefont {Kuraev}\ and\ \citenamefont {Fadin}(1985)}]{Kuraev:1985hb}%
  \BibitemOpen
  \bibfield  {author} {\bibinfo {author} {\bibfnamefont {E.~A.}\ \bibnamefont {Kuraev}}\ and\ \bibinfo {author} {\bibfnamefont {V.~S.}\ \bibnamefont {Fadin}},\ }\bibfield  {title} {\bibinfo {title} {{On Radiative Corrections to $e^+ e^-$ Single Photon Annihilation at High-Energy}},\ }\href@noop {} {\bibfield  {journal} {\bibinfo  {journal} {Sov. J. Nucl. Phys.}\ }\textbf {\bibinfo {volume} {41}},\ \bibinfo {pages} {466} (\bibinfo {year} {1985})}\BibitemShut {NoStop}%
\bibitem [{\citenamefont {Blümlein}\ and\ \citenamefont {Kawamura}(2005)}]{Blumlein:2004bs}%
  \BibitemOpen
  \bibfield  {author} {\bibinfo {author} {\bibfnamefont {J.}~\bibnamefont {Blümlein}}\ and\ \bibinfo {author} {\bibfnamefont {H.}~\bibnamefont {Kawamura}},\ }\bibfield  {title} {\bibinfo {title} {{Universal higher order QED corrections to polarized lepton scattering}},\ }\href {https://doi.org/10.1016/j.nuclphysb.2004.12.001} {\bibfield  {journal} {\bibinfo  {journal} {Nucl. Phys. B}\ }\textbf {\bibinfo {volume} {708}},\ \bibinfo {pages} {467} (\bibinfo {year} {2005})},\ \Eprint {https://arxiv.org/abs/hep-ph/0409289} {arXiv:hep-ph/0409289} \BibitemShut {NoStop}%
\bibitem [{\citenamefont {Bertone}\ \emph {et~al.}(2020)\citenamefont {Bertone}, \citenamefont {Cacciari}, \citenamefont {Frixione},\ and\ \citenamefont {Stagnitto}}]{Bertone:2019hks}%
  \BibitemOpen
  \bibfield  {author} {\bibinfo {author} {\bibfnamefont {V.}~\bibnamefont {Bertone}}, \bibinfo {author} {\bibfnamefont {M.}~\bibnamefont {Cacciari}}, \bibinfo {author} {\bibfnamefont {S.}~\bibnamefont {Frixione}},\ and\ \bibinfo {author} {\bibfnamefont {G.}~\bibnamefont {Stagnitto}},\ }\bibfield  {title} {\bibinfo {title} {{The partonic structure of the electron at the next-to-leading logarithmic accuracy in QED}},\ }\href {https://doi.org/10.1007/JHEP03(2020)135} {\bibfield  {journal} {\bibinfo  {journal} {JHEP}\ }\textbf {\bibinfo {volume} {03}},\ \bibinfo {pages} {135}},\ \bibinfo {note} {[Erratum: JHEP 08, 108 (2022)]},\ \Eprint {https://arxiv.org/abs/1911.12040} {arXiv:1911.12040 [hep-ph]} \BibitemShut {NoStop}%
\bibitem [{\citenamefont {Bertone}\ \emph {et~al.}(2022)\citenamefont {Bertone}, \citenamefont {Cacciari}, \citenamefont {Frixione}, \citenamefont {Stagnitto}, \citenamefont {Zaro},\ and\ \citenamefont {Zhao}}]{Bertone:2022ktl}%
  \BibitemOpen
  \bibfield  {author} {\bibinfo {author} {\bibfnamefont {V.}~\bibnamefont {Bertone}}, \bibinfo {author} {\bibfnamefont {M.}~\bibnamefont {Cacciari}}, \bibinfo {author} {\bibfnamefont {S.}~\bibnamefont {Frixione}}, \bibinfo {author} {\bibfnamefont {G.}~\bibnamefont {Stagnitto}}, \bibinfo {author} {\bibfnamefont {M.}~\bibnamefont {Zaro}},\ and\ \bibinfo {author} {\bibfnamefont {X.}~\bibnamefont {Zhao}},\ }\bibfield  {title} {\bibinfo {title} {{Improving methods and predictions at high-energy $e^{+}e^{-}$ colliders within collinear factorisation}},\ }\href {https://doi.org/10.1007/JHEP10(2022)089} {\bibfield  {journal} {\bibinfo  {journal} {JHEP}\ }\textbf {\bibinfo {volume} {10}},\ \bibinfo {pages} {089}},\ \Eprint {https://arxiv.org/abs/2207.03265} {arXiv:2207.03265 [hep-ph]} \BibitemShut {NoStop}%
\bibitem [{\citenamefont {Schnubel}\ and\ \citenamefont {Szafron}(2025)}]{Schnubel:2025ejl}%
  \BibitemOpen
  \bibfield  {author} {\bibinfo {author} {\bibfnamefont {M.}~\bibnamefont {Schnubel}}\ and\ \bibinfo {author} {\bibfnamefont {R.}~\bibnamefont {Szafron}},\ }\bibfield  {title} {\bibinfo {title} {{Electron and photon structure functions at two loops}},\ }\href {https://doi.org/10.1007/JHEP12(2025)167} {\bibfield  {journal} {\bibinfo  {journal} {JHEP}\ }\textbf {\bibinfo {volume} {12}},\ \bibinfo {pages} {167}},\ \Eprint {https://arxiv.org/abs/2509.09618} {arXiv:2509.09618 [hep-ph]} \BibitemShut {NoStop}%
\bibitem [{\citenamefont {Frixione}\ and\ \citenamefont {Sch{\"o}nwald}(2026)}]{Frixione:2026rii}%
  \BibitemOpen
  \bibfield  {author} {\bibinfo {author} {\bibfnamefont {S.}~\bibnamefont {Frixione}}\ and\ \bibinfo {author} {\bibfnamefont {K.}~\bibnamefont {Sch{\"o}nwald}},\ }\href@noop {} {\bibinfo {title} {{The electron parton distribution functions at the NNLO in QED}}} (\bibinfo {year} {2026}),\ \Eprint {https://arxiv.org/abs/2608.21181} {arXiv:2608.21181 [hep-ph]} \BibitemShut {NoStop}%
\bibitem [{Note3()}]{Note3}%
  \BibitemOpen
  \bibinfo {note} {Note that the use of electron PDFs is not necessary and does not impact the general argument of this Letter.}\BibitemShut {Stop}%
\bibitem [{\citenamefont {Becher}\ \emph {et~al.}(2025{\natexlab{b}})\citenamefont {Becher}, \citenamefont {Hager}, \citenamefont {Jaskiewicz}, \citenamefont {Neubert},\ and\ \citenamefont {Schwienbacher}}]{Becher:2024kmk}%
  \BibitemOpen
  \bibfield  {author} {\bibinfo {author} {\bibfnamefont {T.}~\bibnamefont {Becher}}, \bibinfo {author} {\bibfnamefont {P.}~\bibnamefont {Hager}}, \bibinfo {author} {\bibfnamefont {S.}~\bibnamefont {Jaskiewicz}}, \bibinfo {author} {\bibfnamefont {M.}~\bibnamefont {Neubert}},\ and\ \bibinfo {author} {\bibfnamefont {D.}~\bibnamefont {Schwienbacher}},\ }\bibfield  {title} {\bibinfo {title} {{Factorization Restoration through Glauber Gluons}},\ }\href {https://doi.org/10.1103/PhysRevLett.134.061901} {\bibfield  {journal} {\bibinfo  {journal} {Phys. Rev. Lett.}\ }\textbf {\bibinfo {volume} {134}},\ \bibinfo {pages} {061901} (\bibinfo {year} {2025}{\natexlab{b}})},\ \Eprint {https://arxiv.org/abs/2408.10308} {arXiv:2408.10308 [hep-ph]} \BibitemShut {NoStop}%
\bibitem [{\citenamefont {Becher}\ \emph {et~al.}(2026{\natexlab{d}})\citenamefont {Becher}, \citenamefont {Hager}, \citenamefont {Neubert},\ and\ \citenamefont {Schwienbacher}}]{Becher:2026kbr}%
  \BibitemOpen
  \bibfield  {author} {\bibinfo {author} {\bibfnamefont {T.}~\bibnamefont {Becher}}, \bibinfo {author} {\bibfnamefont {P.}~\bibnamefont {Hager}}, \bibinfo {author} {\bibfnamefont {M.}~\bibnamefont {Neubert}},\ and\ \bibinfo {author} {\bibfnamefont {D.}~\bibnamefont {Schwienbacher}},\ }\bibfield  {title} {\bibinfo {title} {{Factorization beyond Coherence}},\ }\href {https://doi.org/10.1103/q3m4-bcpl} {\bibfield  {journal} {\bibinfo  {journal} {Phys. Rev. Lett.}\ }\textbf {\bibinfo {volume} {137}},\ \bibinfo {pages} {121902} (\bibinfo {year} {2026}{\natexlab{d}})},\ \Eprint {https://arxiv.org/abs/2603.12383} {arXiv:2603.12383 [hep-ph]} \BibitemShut {NoStop}%
\bibitem [{\citenamefont {Rothstein}\ and\ \citenamefont {Stewart}(2016)}]{Rothstein:2016bsq}%
  \BibitemOpen
  \bibfield  {author} {\bibinfo {author} {\bibfnamefont {I.~Z.}\ \bibnamefont {Rothstein}}\ and\ \bibinfo {author} {\bibfnamefont {I.~W.}\ \bibnamefont {Stewart}},\ }\bibfield  {title} {\bibinfo {title} {{An Effective Field Theory for Forward Scattering and Factorization Violation}},\ }\href {https://doi.org/10.1007/JHEP08(2016)025} {\bibfield  {journal} {\bibinfo  {journal} {JHEP}\ }\textbf {\bibinfo {volume} {08}},\ \bibinfo {pages} {025}},\ \Eprint {https://arxiv.org/abs/1601.04695} {arXiv:1601.04695 [hep-ph]} \BibitemShut {NoStop}%
\bibitem [{Note4()}]{Note4}%
  \BibitemOpen
  \bibinfo {note} {This is the scaling for collinear--soft Glauber, it is also possible to have anti-collinear--collinear Glaubers as well as anti-collinear--soft Glaubers which differ in scaling. Notably, the transverse momentum has to be compatible with both directions it mediates in between.}\BibitemShut {Stop}%
\bibitem [{\citenamefont {Han}\ \emph {et~al.}(2022)\citenamefont {Han}, \citenamefont {Ma},\ and\ \citenamefont {Xie}}]{Han:2021kes}%
  \BibitemOpen
  \bibfield  {author} {\bibinfo {author} {\bibfnamefont {T.}~\bibnamefont {Han}}, \bibinfo {author} {\bibfnamefont {Y.}~\bibnamefont {Ma}},\ and\ \bibinfo {author} {\bibfnamefont {K.}~\bibnamefont {Xie}},\ }\bibfield  {title} {\bibinfo {title} {{Quark and gluon contents of a lepton at high energies}},\ }\href {https://doi.org/10.1007/JHEP02(2022)154} {\bibfield  {journal} {\bibinfo  {journal} {JHEP}\ }\textbf {\bibinfo {volume} {02}},\ \bibinfo {pages} {154}},\ \Eprint {https://arxiv.org/abs/2103.09844} {arXiv:2103.09844 [hep-ph]} \BibitemShut {NoStop}%
\bibitem [{\citenamefont {Garosi}\ \emph {et~al.}(2023)\citenamefont {Garosi}, \citenamefont {Marzocca},\ and\ \citenamefont {Trifinopoulos}}]{Garosi:2023bvq}%
  \BibitemOpen
  \bibfield  {author} {\bibinfo {author} {\bibfnamefont {F.}~\bibnamefont {Garosi}}, \bibinfo {author} {\bibfnamefont {D.}~\bibnamefont {Marzocca}},\ and\ \bibinfo {author} {\bibfnamefont {S.}~\bibnamefont {Trifinopoulos}},\ }\bibfield  {title} {\bibinfo {title} {{LePDF: Standard Model PDFs for high-energy lepton colliders}},\ }\href {https://doi.org/10.1007/JHEP09(2023)107} {\bibfield  {journal} {\bibinfo  {journal} {JHEP}\ }\textbf {\bibinfo {volume} {09}},\ \bibinfo {pages} {107}},\ \Eprint {https://arxiv.org/abs/2303.16964} {arXiv:2303.16964 [hep-ph]} \BibitemShut {NoStop}%
\bibitem [{\citenamefont {Bodwin}(1985)}]{Bodwin:1984hc}%
  \BibitemOpen
  \bibfield  {author} {\bibinfo {author} {\bibfnamefont {G.~T.}\ \bibnamefont {Bodwin}},\ }\bibfield  {title} {\bibinfo {title} {{Factorization of the Drell-Yan Cross-Section in Perturbation Theory}},\ }\href {https://doi.org/10.1103/PhysRevD.34.3932} {\bibfield  {journal} {\bibinfo  {journal} {Phys. Rev. D}\ }\textbf {\bibinfo {volume} {31}},\ \bibinfo {pages} {2616} (\bibinfo {year} {1985})},\ \bibinfo {note} {[Erratum: Phys.Rev.D 34, 3932 (1986)]}\BibitemShut {NoStop}%
\bibitem [{\citenamefont {Aybat}\ and\ \citenamefont {Sterman}(2009)}]{Aybat:2008ct}%
  \BibitemOpen
  \bibfield  {author} {\bibinfo {author} {\bibfnamefont {S.~M.}\ \bibnamefont {Aybat}}\ and\ \bibinfo {author} {\bibfnamefont {G.~F.}\ \bibnamefont {Sterman}},\ }\bibfield  {title} {\bibinfo {title} {{Soft-Gluon Cancellation, Phases and Factorization with Initial-State Partons}},\ }\href {https://doi.org/10.1016/j.physletb.2008.11.050} {\bibfield  {journal} {\bibinfo  {journal} {Phys. Lett. B}\ }\textbf {\bibinfo {volume} {671}},\ \bibinfo {pages} {46} (\bibinfo {year} {2009})},\ \Eprint {https://arxiv.org/abs/0811.0246} {arXiv:0811.0246 [hep-ph]} \BibitemShut {NoStop}%
\bibitem [{\citenamefont {Sterman}(2022)}]{Sterman:2022gyf}%
  \BibitemOpen
  \bibfield  {author} {\bibinfo {author} {\bibfnamefont {G.}~\bibnamefont {Sterman}},\ }\bibfield  {title} {\bibinfo {title} {{Comments on collinear factorization}},\ }in\ \href@noop {} {\emph {\bibinfo {booktitle} {{Snowmass 2021}}}}\ (\bibinfo {year} {2022})\ \Eprint {https://arxiv.org/abs/2207.06507} {arXiv:2207.06507 [hep-ph]} \BibitemShut {NoStop}%
\bibitem [{\citenamefont {Sterman}\ and\ \citenamefont {Yu}(2026)}]{Sterman:2026vuh}%
  \BibitemOpen
  \bibfield  {author} {\bibinfo {author} {\bibfnamefont {G.}~\bibnamefont {Sterman}}\ and\ \bibinfo {author} {\bibfnamefont {Z.}~\bibnamefont {Yu}},\ }\href@noop {} {\bibinfo {title} {{Collinear Factorization and Evolution}}} (\bibinfo {year} {2026}),\ \Eprint {https://arxiv.org/abs/2610.38666} {arXiv:2610.38666 [hep-ph]} \BibitemShut {NoStop}%
\bibitem [{\citenamefont {Banfi}\ \emph {et~al.}(2026)\citenamefont {Banfi}, \citenamefont {Forshaw},\ and\ \citenamefont {Holguin}}]{Banfi:2025mra}%
  \BibitemOpen
  \bibfield  {author} {\bibinfo {author} {\bibfnamefont {A.}~\bibnamefont {Banfi}}, \bibinfo {author} {\bibfnamefont {J.~R.}\ \bibnamefont {Forshaw}},\ and\ \bibinfo {author} {\bibfnamefont {J.}~\bibnamefont {Holguin}},\ }\bibfield  {title} {\bibinfo {title} {{One-Jettiness Distribution Contains Super-Super-Leading Logarithms}},\ }\href {https://doi.org/10.1103/flv7-ksvf} {\bibfield  {journal} {\bibinfo  {journal} {Phys. Rev. Lett.}\ }\textbf {\bibinfo {volume} {136}},\ \bibinfo {pages} {221901} (\bibinfo {year} {2026})},\ \Eprint {https://arxiv.org/abs/2511.11799} {arXiv:2511.11799 [hep-ph]} \BibitemShut {NoStop}%
\bibitem [{\citenamefont {Forshaw}\ \emph {et~al.}(2012)\citenamefont {Forshaw}, \citenamefont {Seymour},\ and\ \citenamefont {Siodmok}}]{Forshaw:2012bi}%
  \BibitemOpen
  \bibfield  {author} {\bibinfo {author} {\bibfnamefont {J.~R.}\ \bibnamefont {Forshaw}}, \bibinfo {author} {\bibfnamefont {M.~H.}\ \bibnamefont {Seymour}},\ and\ \bibinfo {author} {\bibfnamefont {A.}~\bibnamefont {Siodmok}},\ }\bibfield  {title} {\bibinfo {title} {{On the Breaking of Collinear Factorization in QCD}},\ }\href {https://doi.org/10.1007/JHEP11(2012)066} {\bibfield  {journal} {\bibinfo  {journal} {JHEP}\ }\textbf {\bibinfo {volume} {11}},\ \bibinfo {pages} {066}},\ \Eprint {https://arxiv.org/abs/1206.6363} {arXiv:1206.6363 [hep-ph]} \BibitemShut {NoStop}%
\bibitem [{\citenamefont {Stewart}\ \emph {et~al.}(2010)\citenamefont {Stewart}, \citenamefont {Tackmann},\ and\ \citenamefont {Waalewijn}}]{Stewart:2010tn}%
  \BibitemOpen
  \bibfield  {author} {\bibinfo {author} {\bibfnamefont {I.~W.}\ \bibnamefont {Stewart}}, \bibinfo {author} {\bibfnamefont {F.~J.}\ \bibnamefont {Tackmann}},\ and\ \bibinfo {author} {\bibfnamefont {W.~J.}\ \bibnamefont {Waalewijn}},\ }\bibfield  {title} {\bibinfo {title} {{N-Jettiness: An Inclusive Event Shape to Veto Jets}},\ }\href {https://doi.org/10.1103/PhysRevLett.105.092002} {\bibfield  {journal} {\bibinfo  {journal} {Phys. Rev. Lett.}\ }\textbf {\bibinfo {volume} {105}},\ \bibinfo {pages} {092002} (\bibinfo {year} {2010})},\ \Eprint {https://arxiv.org/abs/1004.2489} {arXiv:1004.2489 [hep-ph]} \BibitemShut {NoStop}%
\bibitem [{\citenamefont {Dixon}\ \emph {et~al.}(2020)\citenamefont {Dixon}, \citenamefont {Herrmann}, \citenamefont {Yan},\ and\ \citenamefont {Zhu}}]{Dixon:2019lnw}%
  \BibitemOpen
  \bibfield  {author} {\bibinfo {author} {\bibfnamefont {L.~J.}\ \bibnamefont {Dixon}}, \bibinfo {author} {\bibfnamefont {E.}~\bibnamefont {Herrmann}}, \bibinfo {author} {\bibfnamefont {K.}~\bibnamefont {Yan}},\ and\ \bibinfo {author} {\bibfnamefont {H.~X.}\ \bibnamefont {Zhu}},\ }\bibfield  {title} {\bibinfo {title} {{Soft gluon emission at two loops in full color}},\ }\href {https://doi.org/10.1007/JHEP05(2020)135} {\bibfield  {journal} {\bibinfo  {journal} {JHEP}\ }\textbf {\bibinfo {volume} {05}},\ \bibinfo {pages} {135}},\ \bibinfo {note} {[Erratum: JHEP 06, 143 (2024)]},\ \Eprint {https://arxiv.org/abs/1912.09370} {arXiv:1912.09370 [hep-ph]} \BibitemShut {NoStop}%
\bibitem [{\citenamefont {Gaunt}(2014)}]{Gaunt:2014ska}%
  \BibitemOpen
  \bibfield  {author} {\bibinfo {author} {\bibfnamefont {J.~R.}\ \bibnamefont {Gaunt}},\ }\bibfield  {title} {\bibinfo {title} {{Glauber Gluons and Multiple Parton Interactions}},\ }\href {https://doi.org/10.1007/JHEP07(2014)110} {\bibfield  {journal} {\bibinfo  {journal} {JHEP}\ }\textbf {\bibinfo {volume} {07}},\ \bibinfo {pages} {110}},\ \Eprint {https://arxiv.org/abs/1405.2080} {arXiv:1405.2080 [hep-ph]} \BibitemShut {NoStop}%
\bibitem [{\citenamefont {Zeng}(2015)}]{Zeng:2015iba}%
  \BibitemOpen
  \bibfield  {author} {\bibinfo {author} {\bibfnamefont {M.}~\bibnamefont {Zeng}},\ }\bibfield  {title} {\bibinfo {title} {{Drell-Yan process with jet vetoes: breaking of generalized factorization}},\ }\href {https://doi.org/10.1007/JHEP10(2015)189} {\bibfield  {journal} {\bibinfo  {journal} {JHEP}\ }\textbf {\bibinfo {volume} {10}},\ \bibinfo {pages} {189}},\ \Eprint {https://arxiv.org/abs/1507.01652} {arXiv:1507.01652 [hep-ph]} \BibitemShut {NoStop}%
\bibitem [{\citenamefont {Collins}\ \emph {et~al.}(1993)\citenamefont {Collins}, \citenamefont {Frankfurt},\ and\ \citenamefont {Strikman}}]{Collins:1992cv}%
  \BibitemOpen
  \bibfield  {author} {\bibinfo {author} {\bibfnamefont {J.~C.}\ \bibnamefont {Collins}}, \bibinfo {author} {\bibfnamefont {L.}~\bibnamefont {Frankfurt}},\ and\ \bibinfo {author} {\bibfnamefont {M.}~\bibnamefont {Strikman}},\ }\bibfield  {title} {\bibinfo {title} {{Diffractive hard scattering with a coherent pomeron}},\ }\href {https://doi.org/10.1016/0370-2693(93)90206-W} {\bibfield  {journal} {\bibinfo  {journal} {Phys. Lett. B}\ }\textbf {\bibinfo {volume} {307}},\ \bibinfo {pages} {161} (\bibinfo {year} {1993})},\ \Eprint {https://arxiv.org/abs/hep-ph/9212212} {arXiv:hep-ph/9212212} \BibitemShut {NoStop}%
\bibitem [{\citenamefont {Fichet}\ \emph {et~al.}(2015)\citenamefont {Fichet}, \citenamefont {von Gersdorff}, \citenamefont {Lenzi}, \citenamefont {Royon},\ and\ \citenamefont {Saimpert}}]{Fichet:2014uka}%
  \BibitemOpen
  \bibfield  {author} {\bibinfo {author} {\bibfnamefont {S.}~\bibnamefont {Fichet}}, \bibinfo {author} {\bibfnamefont {G.}~\bibnamefont {von Gersdorff}}, \bibinfo {author} {\bibfnamefont {B.}~\bibnamefont {Lenzi}}, \bibinfo {author} {\bibfnamefont {C.}~\bibnamefont {Royon}},\ and\ \bibinfo {author} {\bibfnamefont {M.}~\bibnamefont {Saimpert}},\ }\bibfield  {title} {\bibinfo {title} {{Light-by-light scattering with intact protons at the LHC: from Standard Model to New Physics}},\ }\href {https://doi.org/10.1007/JHEP02(2015)165} {\bibfield  {journal} {\bibinfo  {journal} {JHEP}\ }\textbf {\bibinfo {volume} {02}},\ \bibinfo {pages} {165}},\ \Eprint {https://arxiv.org/abs/1411.6629} {arXiv:1411.6629 [hep-ph]} \BibitemShut {NoStop}%
\bibitem [{\citenamefont {Frankfurt}\ \emph {et~al.}(2022)\citenamefont {Frankfurt}, \citenamefont {Guzey}, \citenamefont {Stasto},\ and\ \citenamefont {Strikman}}]{Frankfurt:2022jns}%
  \BibitemOpen
  \bibfield  {author} {\bibinfo {author} {\bibfnamefont {L.}~\bibnamefont {Frankfurt}}, \bibinfo {author} {\bibfnamefont {V.}~\bibnamefont {Guzey}}, \bibinfo {author} {\bibfnamefont {A.}~\bibnamefont {Stasto}},\ and\ \bibinfo {author} {\bibfnamefont {M.}~\bibnamefont {Strikman}},\ }\bibfield  {title} {\bibinfo {title} {{Selected topics in diffraction with protons and nuclei: past, present, and future}},\ }\href {https://doi.org/10.1088/1361-6633/ac8228} {\bibfield  {journal} {\bibinfo  {journal} {Rept. Prog. Phys.}\ }\textbf {\bibinfo {volume} {85}},\ \bibinfo {pages} {126301} (\bibinfo {year} {2022})},\ \Eprint {https://arxiv.org/abs/2203.12289} {arXiv:2203.12289 [hep-ph]} \BibitemShut {NoStop}%
\bibitem [{\citenamefont {Lee}\ \emph {et~al.}(2025)\citenamefont {Lee}, \citenamefont {Schindler},\ and\ \citenamefont {Stewart}}]{Lee:2025fml}%
  \BibitemOpen
  \bibfield  {author} {\bibinfo {author} {\bibfnamefont {K.}~\bibnamefont {Lee}}, \bibinfo {author} {\bibfnamefont {S.~T.}\ \bibnamefont {Schindler}},\ and\ \bibinfo {author} {\bibfnamefont {I.~W.}\ \bibnamefont {Stewart}},\ }\bibfield  {title} {\bibinfo {title} {{Effective field theory factorization for diffraction}},\ }\href {https://doi.org/10.1007/JHEP11(2025)157} {\bibfield  {journal} {\bibinfo  {journal} {JHEP}\ }\textbf {\bibinfo {volume} {11}},\ \bibinfo {pages} {157}},\ \Eprint {https://arxiv.org/abs/2508.10231} {arXiv:2508.10231 [hep-ph]} \BibitemShut {NoStop}%
\bibitem [{\citenamefont {Aretz}\ \emph {et~al.}(2026)\citenamefont {Aretz}, \citenamefont {Lee}, \citenamefont {Schindler},\ and\ \citenamefont {Stewart}}]{Aretz:2026sjb}%
  \BibitemOpen
  \bibfield  {author} {\bibinfo {author} {\bibfnamefont {P.~B.}\ \bibnamefont {Aretz}}, \bibinfo {author} {\bibfnamefont {K.}~\bibnamefont {Lee}}, \bibinfo {author} {\bibfnamefont {S.~T.}\ \bibnamefont {Schindler}},\ and\ \bibinfo {author} {\bibfnamefont {I.~W.}\ \bibnamefont {Stewart}},\ }\href@noop {} {\bibinfo {title} {{Factorization of elastic, single, and double diffractive $pp$ scattering}}} (\bibinfo {year} {2026}),\ \Eprint {https://arxiv.org/abs/2607.07788} {arXiv:2607.07788 [hep-ph]} \BibitemShut {NoStop}%
\bibitem [{\citenamefont {Caola}\ \emph {et~al.}(2022)\citenamefont {Caola}, \citenamefont {Chakraborty}, \citenamefont {Gambuti}, \citenamefont {von Manteuffel},\ and\ \citenamefont {Tancredi}}]{Caola:2021izf}%
  \BibitemOpen
  \bibfield  {author} {\bibinfo {author} {\bibfnamefont {F.}~\bibnamefont {Caola}}, \bibinfo {author} {\bibfnamefont {A.}~\bibnamefont {Chakraborty}}, \bibinfo {author} {\bibfnamefont {G.}~\bibnamefont {Gambuti}}, \bibinfo {author} {\bibfnamefont {A.}~\bibnamefont {von Manteuffel}},\ and\ \bibinfo {author} {\bibfnamefont {L.}~\bibnamefont {Tancredi}},\ }\bibfield  {title} {\bibinfo {title} {{Three-Loop Gluon Scattering in QCD and the Gluon Regge Trajectory}},\ }\href {https://doi.org/10.1103/PhysRevLett.128.212001} {\bibfield  {journal} {\bibinfo  {journal} {Phys. Rev. Lett.}\ }\textbf {\bibinfo {volume} {128}},\ \bibinfo {pages} {212001} (\bibinfo {year} {2022})},\ \Eprint {https://arxiv.org/abs/2112.11097} {arXiv:2112.11097 [hep-ph]} \BibitemShut {NoStop}%
\bibitem [{\citenamefont {Falcioni}\ \emph {et~al.}(2022{\natexlab{a}})\citenamefont {Falcioni}, \citenamefont {Gardi}, \citenamefont {Maher}, \citenamefont {Milloy},\ and\ \citenamefont {Vernazza}}]{Falcioni:2021buo}%
  \BibitemOpen
  \bibfield  {author} {\bibinfo {author} {\bibfnamefont {G.}~\bibnamefont {Falcioni}}, \bibinfo {author} {\bibfnamefont {E.}~\bibnamefont {Gardi}}, \bibinfo {author} {\bibfnamefont {N.}~\bibnamefont {Maher}}, \bibinfo {author} {\bibfnamefont {C.}~\bibnamefont {Milloy}},\ and\ \bibinfo {author} {\bibfnamefont {L.}~\bibnamefont {Vernazza}},\ }\bibfield  {title} {\bibinfo {title} {{Scattering amplitudes in the Regge limit and the soft anomalous dimension through four loops}},\ }\href {https://doi.org/10.1007/JHEP03(2022)053} {\bibfield  {journal} {\bibinfo  {journal} {JHEP}\ }\textbf {\bibinfo {volume} {03}},\ \bibinfo {pages} {053}},\ \Eprint {https://arxiv.org/abs/2111.10664} {arXiv:2111.10664 [hep-ph]} \BibitemShut {NoStop}%
\bibitem [{\citenamefont {Falcioni}\ \emph {et~al.}(2022{\natexlab{b}})\citenamefont {Falcioni}, \citenamefont {Gardi}, \citenamefont {Maher}, \citenamefont {Milloy},\ and\ \citenamefont {Vernazza}}]{Falcioni:2021dgr}%
  \BibitemOpen
  \bibfield  {author} {\bibinfo {author} {\bibfnamefont {G.}~\bibnamefont {Falcioni}}, \bibinfo {author} {\bibfnamefont {E.}~\bibnamefont {Gardi}}, \bibinfo {author} {\bibfnamefont {N.}~\bibnamefont {Maher}}, \bibinfo {author} {\bibfnamefont {C.}~\bibnamefont {Milloy}},\ and\ \bibinfo {author} {\bibfnamefont {L.}~\bibnamefont {Vernazza}},\ }\bibfield  {title} {\bibinfo {title} {{Disentangling the Regge Cut and Regge Pole in Perturbative QCD}},\ }\href {https://doi.org/10.1103/PhysRevLett.128.132001} {\bibfield  {journal} {\bibinfo  {journal} {Phys. Rev. Lett.}\ }\textbf {\bibinfo {volume} {128}},\ \bibinfo {pages} {132001} (\bibinfo {year} {2022}{\natexlab{b}})},\ \Eprint {https://arxiv.org/abs/2112.11098} {arXiv:2112.11098 [hep-ph]} \BibitemShut {NoStop}%
\bibitem [{\citenamefont {Neill}\ \emph {et~al.}(2023)\citenamefont {Neill}, \citenamefont {Pathak},\ and\ \citenamefont {Stewart}}]{Neill:2023jcd}%
  \BibitemOpen
  \bibfield  {author} {\bibinfo {author} {\bibfnamefont {D.}~\bibnamefont {Neill}}, \bibinfo {author} {\bibfnamefont {A.}~\bibnamefont {Pathak}},\ and\ \bibinfo {author} {\bibfnamefont {I.~W.}\ \bibnamefont {Stewart}},\ }\bibfield  {title} {\bibinfo {title} {{Small-$x$ factorization from effective field theory}},\ }\href {https://doi.org/10.1007/JHEP09(2023)089} {\bibfield  {journal} {\bibinfo  {journal} {JHEP}\ }\textbf {\bibinfo {volume} {09}},\ \bibinfo {pages} {089}},\ \Eprint {https://arxiv.org/abs/2303.13710} {arXiv:2303.13710 [hep-ph]} \BibitemShut {NoStop}%
\bibitem [{\citenamefont {Gao}\ \emph {et~al.}(2024{\natexlab{a}})\citenamefont {Gao}, \citenamefont {Moult}, \citenamefont {Raman}, \citenamefont {Ridgway},\ and\ \citenamefont {Stewart}}]{Gao:2024qsg}%
  \BibitemOpen
  \bibfield  {author} {\bibinfo {author} {\bibfnamefont {A.}~\bibnamefont {Gao}}, \bibinfo {author} {\bibfnamefont {I.}~\bibnamefont {Moult}}, \bibinfo {author} {\bibfnamefont {S.}~\bibnamefont {Raman}}, \bibinfo {author} {\bibfnamefont {G.}~\bibnamefont {Ridgway}},\ and\ \bibinfo {author} {\bibfnamefont {I.~W.}\ \bibnamefont {Stewart}},\ }\bibfield  {title} {\bibinfo {title} {{A collinear perspective on the Regge limit}},\ }\href {https://doi.org/10.1007/JHEP05(2024)328} {\bibfield  {journal} {\bibinfo  {journal} {JHEP}\ }\textbf {\bibinfo {volume} {05}},\ \bibinfo {pages} {328}},\ \Eprint {https://arxiv.org/abs/2401.00931} {arXiv:2401.00931 [hep-ph]} \BibitemShut {NoStop}%
\bibitem [{\citenamefont {Buccioni}\ \emph {et~al.}(2025)\citenamefont {Buccioni}, \citenamefont {Caola}, \citenamefont {Devoto},\ and\ \citenamefont {Gambuti}}]{Buccioni:2024gzo}%
  \BibitemOpen
  \bibfield  {author} {\bibinfo {author} {\bibfnamefont {F.}~\bibnamefont {Buccioni}}, \bibinfo {author} {\bibfnamefont {F.}~\bibnamefont {Caola}}, \bibinfo {author} {\bibfnamefont {F.}~\bibnamefont {Devoto}},\ and\ \bibinfo {author} {\bibfnamefont {G.}~\bibnamefont {Gambuti}},\ }\bibfield  {title} {\bibinfo {title} {{Investigating the universality of five-point QCD scattering amplitudes at high energy}},\ }\href {https://doi.org/10.1007/JHEP03(2025)129} {\bibfield  {journal} {\bibinfo  {journal} {JHEP}\ }\textbf {\bibinfo {volume} {03}},\ \bibinfo {pages} {129}},\ \Eprint {https://arxiv.org/abs/2411.14050} {arXiv:2411.14050 [hep-ph]} \BibitemShut {NoStop}%
\bibitem [{\citenamefont {Gao}\ \emph {et~al.}(2024{\natexlab{b}})\citenamefont {Gao}, \citenamefont {Moult}, \citenamefont {Raman}, \citenamefont {Ridgway},\ and\ \citenamefont {Stewart}}]{Gao:2024fyz}%
  \BibitemOpen
  \bibfield  {author} {\bibinfo {author} {\bibfnamefont {A.}~\bibnamefont {Gao}}, \bibinfo {author} {\bibfnamefont {I.}~\bibnamefont {Moult}}, \bibinfo {author} {\bibfnamefont {S.}~\bibnamefont {Raman}}, \bibinfo {author} {\bibfnamefont {G.}~\bibnamefont {Ridgway}},\ and\ \bibinfo {author} {\bibfnamefont {I.~W.}\ \bibnamefont {Stewart}},\ }\href@noop {} {\bibinfo {title} {{Reggeization in Color}}} (\bibinfo {year} {2024}{\natexlab{b}}),\ \Eprint {https://arxiv.org/abs/2411.09692} {arXiv:2411.09692 [hep-ph]} \BibitemShut {NoStop}%
\bibitem [{\citenamefont {Barcaro}\ \emph {et~al.}(2026)\citenamefont {Barcaro}, \citenamefont {Gao}, \citenamefont {Gaunt},\ and\ \citenamefont {Pathak}}]{Barcaro:2026dsd}%
  \BibitemOpen
  \bibfield  {author} {\bibinfo {author} {\bibfnamefont {D.}~\bibnamefont {Barcaro}}, \bibinfo {author} {\bibfnamefont {A.}~\bibnamefont {Gao}}, \bibinfo {author} {\bibfnamefont {J.~R.}\ \bibnamefont {Gaunt}},\ and\ \bibinfo {author} {\bibfnamefont {A.}~\bibnamefont {Pathak}},\ }\href@noop {} {\bibinfo {title} {{Collinear Factorization Violation and Reggeization}}} (\bibinfo {year} {2026}),\ \Eprint {https://arxiv.org/abs/2607.14259} {arXiv:2607.14259 [hep-ph]} \BibitemShut {NoStop}%
\bibitem [{\citenamefont {Davies}\ \emph {et~al.}(2022)\citenamefont {Davies}, \citenamefont {Kom}, \citenamefont {Moch},\ and\ \citenamefont {Vogt}}]{Davies:2022ofz}%
  \BibitemOpen
  \bibfield  {author} {\bibinfo {author} {\bibfnamefont {J.}~\bibnamefont {Davies}}, \bibinfo {author} {\bibfnamefont {C.~H.}\ \bibnamefont {Kom}}, \bibinfo {author} {\bibfnamefont {S.}~\bibnamefont {Moch}},\ and\ \bibinfo {author} {\bibfnamefont {A.}~\bibnamefont {Vogt}},\ }\bibfield  {title} {\bibinfo {title} {{Resummation of small-$x$ double logarithms in QCD: inclusive deep-inelastic scattering}},\ }\href {https://doi.org/10.1007/JHEP08(2022)135} {\bibfield  {journal} {\bibinfo  {journal} {JHEP}\ }\textbf {\bibinfo {volume} {08}},\ \bibinfo {pages} {135}},\ \Eprint {https://arxiv.org/abs/2202.10362} {arXiv:2202.10362 [hep-ph]} \BibitemShut {NoStop}%
\bibitem [{\citenamefont {Gehrmann}\ \emph {et~al.}(2026)\citenamefont {Gehrmann}, \citenamefont {von Manteuffel}, \citenamefont {Sotnikov},\ and\ \citenamefont {Yang}}]{Gehrmann:2026qbl}%
  \BibitemOpen
  \bibfield  {author} {\bibinfo {author} {\bibfnamefont {T.}~\bibnamefont {Gehrmann}}, \bibinfo {author} {\bibfnamefont {A.}~\bibnamefont {von Manteuffel}}, \bibinfo {author} {\bibfnamefont {V.}~\bibnamefont {Sotnikov}},\ and\ \bibinfo {author} {\bibfnamefont {T.-Z.}\ \bibnamefont {Yang}},\ }\href@noop {} {\bibinfo {title} {{The four-loop non-singlet splitting functions in QCD}}} (\bibinfo {year} {2026}),\ \Eprint {https://arxiv.org/abs/2604.09534} {arXiv:2604.09534 [hep-ph]} \BibitemShut {NoStop}%
\bibitem [{\citenamefont {Moch}\ and\ \citenamefont {Vogt}(2026)}]{Moch:2026qsw}%
  \BibitemOpen
  \bibfield  {author} {\bibinfo {author} {\bibfnamefont {S.}~\bibnamefont {Moch}}\ and\ \bibinfo {author} {\bibfnamefont {A.}~\bibnamefont {Vogt}},\ }\bibfield  {title} {\bibinfo {title} {{Properties and implications of the four-loop non-singlet splitting functions in QCD}},\ }\href {https://doi.org/10.1007/JHEP09(2026)227} {\bibfield  {journal} {\bibinfo  {journal} {JHEP}\ }\textbf {\bibinfo {volume} {09}},\ \bibinfo {pages} {227}},\ \Eprint {https://arxiv.org/abs/2605.03889} {arXiv:2605.03889 [hep-ph]} \BibitemShut {NoStop}%
\end{thebibliography}%

\end{document}